\documentclass[12pt]{article}
\usepackage{a4wide,amssymb,cite}
\pdfoutput=1

\usepackage{a4wide,amssymb,graphicx}
\usepackage{epsfig}
\usepackage[usenames,dvipsnames]{color}
\usepackage{slashed}
\usepackage{amssymb,cite,graphicx}
\usepackage{slashed}
\usepackage{amsmath,bm,bbm}
\usepackage{amsfonts}
\usepackage[titletoc,title]{appendix}
\usepackage[small]{caption}
\usepackage[margin=1in]{geometry}
\usepackage[multiple]{footmisc}
\usepackage{mathtools}
\usepackage{slashed}
\usepackage[nottoc]{tocbibind}
\usepackage{xcolor}

\newcommand{\be}{\begin{equation}}
\newcommand{\ee}{\end{equation}}
\newcommand{\bea}{\begin{eqnarray}}
\newcommand{\eea}{\end{eqnarray}}

\def\circa#1{\,\raise.3ex\hbox{$#1$\kern-.75em\lower1ex\hbox{$\sim$}}\,}

\begin{document}

\begin{titlepage}

\rightline{CERN-TH-2020-033}

\begin{centering}
\vspace{1cm}
{\Large {\bf Effective theory for self-interacting dark matter \vspace{.2cm} \\ and massive spin-2 mediators}} \\

\vspace{1.5cm}

{\bf  Yoo-Jin Kang$^\star$ and Hyun Min Lee$^\dagger$ }
\vspace{.5cm}

{\it Department of Physics, Chung-Ang University, Seoul 06974, Korea.} 
\\ \vspace{0.2cm}
{\it CERN, Theory department, 1211 Geneva 23, Switzerland. }

\end{centering}
\vspace{2cm}

\begin{abstract}
\noindent
We consider the effective theory for self-interacting dark matter with arbitrary spin and go beyond the previous discussion in the literature by introducing a massive spin-2 particle as the mediator for the dark matter self-scattering. 
We present the effective self-interactions for dark matter in the leading order expansions with the momentum transfer and the dark matter velocity.    
We compare between the Born cross section and  the non-perturbative cross section in the leading order approximation of the effective Yukawa interaction. As a result, we find that there is a wide range of dark matter and spin-2 particle masses for a velocity-dependent self-scattering to solve small-scale problems at galaxies as well as satisfy the bounds from galaxy clusters at the same time.

\end{abstract}

\vspace{3cm}

 \begin{flushleft} 
 $^\star$Email: yoojinkang91@gmail.com \\
$^\dagger$Email: hminlee@cau.ac.kr 
 \end{flushleft}

\end{titlepage}

\section{Introduction}

Collision-less and cold dark matter (DM) explains large scale structures well at about Mpc scale and beyond, but there have been discrepancies between the structure formation obtained from $N$-body simulations and the observed rotation curves in galaxies.  The too-big-to-fail problem has to do with the fact that the observed subhalos in galaxies are less massive than those in the $N$-body simulations, thus lead to the small-scale problems together with the core-cusp problem \cite{small-scale,small-scale2}. This becomes a motivation for us to go beyond the Weakly Interacting Massive Particles (WIMP) paradigm. 

Recently, self-interacting dark matter (SIDM) has drawn a lot of attention from both astrophysics and particle physics communities, because it gives rise to a resolution to the small-scale problems, provided that the self-scattering cross section per DM mass is given in the range of $\sigma_{\rm self}/m_{\rm DM}=0.1-10\,{\rm cm^2/g}$ \cite{Yu}.  
The effects of baryons and supernova feedback in simulations \cite{baryon} might resolve small-scale problems in massive galaxies but it would be also worthwhile to investigate alternatives to WIMPs such as dark matter with new long-range forces and/or light dark matter below GeV scale, that are consistent with the bounds from galaxy clusters \cite{bullet}.

For the purpose of making the discussion on dark matter  as general as possible, we can take the effective theory approach where a set of effective operators are introduced being consistent with the symmetries given at energy scales.
However, the coefficients of the effective operators are often correlated between one and another in a concrete microscopic model for dark matter. In particular, given that there is no evidence for dark matter to carry the charges under the gauge symmetries of the Standard Model (SM), it is legitimate to assume that dark matter is neutral but constrain the properties and interactions of dark matter by the mediator particles. Depending on the spins of mediators and dark matter, there are a handful number of consistent effective operators for dark matter to capture the model dependences and compare with experimental data.

In this article, taking the effective theory approach for the self-interactions of dark matter, we extend the Parity-invariant effective potential for self-interacting dark matter in the literature \cite{tanedo,reece} by including the massive spin-2 particle as the mediator for dark matter \cite{GMDM1,GMDM2,diphoton,cascade}.  Assuming that dark matter carries spin $s=0,1/2$ or $1$ and it couples to the massive spin-2 particle via the energy-momentum tensor, we show the general consequences for the effective self-interactions of dark matter in each case. When the contributions from other mediators such as spin-0 and spin-1 particles are present, the results from the massive spin-2 particle only should be generalized. But, we assume the dominance of the massive spin-2 particle as a working hypothesis and show the important features of the resulting effective self-interactions for dark matter.

The massive spin-2 particle was not taken into consideration in Ref.~\cite{tanedo} due to the lack of the derivative expansion at scales comparable to the mass of the spin-2 mediator. But, in our case, we keep the dark matter momentum and the momentum transfer to be below the mass of the spin-2 mediator in the effective theory for dark matter and discuss the salient feature of the dark matter self-scattering in the Born and quantum regimes. Therefore, there is no contradiction with Ref.~\cite{tanedo}. 

There have been already a lot of papers in the literature, related to the massive spin-2 mediator in connection to WIMPs, light dark matter, as well as Feebly Interacting Massive Particles  \cite{GMDM1,GMDM2,diphoton,cascade, GMetc,DD,GLDM}. In this case, the linear interactions of the massive spin-2 mediator to the energy-momentum tensors for the SM and dark matter have been considered. However, there was no systematic study in the literature discussing the role of the massive spin-2 mediator in the dark matter self-scattering in the effective theory, other than in the Born approximation in the paper of the current authors \cite{GLDM}.

We first consider the non-relativistic effective theory for dark matter and the massive spin-2 mediator and identify the coefficients of the effective operators for self-scattering in the leading order expansions with the momentum transfer and the DM velocity. 
Then, we show explicitly the dependences of our results on the mass and spin of dark matter as well as the mediator mass. 
We take the effective Yukawa potential to be dominant in the perturbative expansion and calculate the Born cross section and the non-perturbative cross section for dark matter self-scattering as well as the Sommerfeld factor for dark matter annihilation. It would be desirable to make a further analysis  with spin-dependent interactions  in more detail, but we leave it for a future study.

We present the most general effective potentials for dark matter beyond the spin-0 and spin-1 mediators discussed in the literature \cite{tanedo,reece}. Thus, we make a further development for the effective field theory with the leading effective Yukawa potential by including the spin-dependent interactions such as spin-spin interactions, dipole interactions, spin-orbit couplings, etc. Moreover, the leading Yukawa-type interactions have the dependences on the mass of the spin-2 particle, due to the fact that the spin-2 mediator has derivative couplings from the energy-momentum tensor.
Moreover, while the massive spin-2 mediator induces similar spin-dependent interactions for fermion dark matter as in the cases of spin-0 and spin-1 mediators, there are new types of spin-dependent interactions for vector dark matter because of the extra polarization states of the spin-2 particle.  

In the case of a light spin-2 particle, we show that the Born approximation and the non-perturbative effects give rise to a velocity-dependent self-scattering for dark matter. We also address the Sommerfeld effects for dark matter annihilation and tie them to the non-perturbative effects for dark matter self-scattering such as bound-state effects.  As a consequence, for appropriate choices of the coupling and mass of the spin-2 mediator, we can achieve a large self-scattering cross section for dark matter to solve the small-scale problems at galaxy scales, while being in the validity of momentum expansion for the massive spin-2 mediator and satisfying the limits from galaxy clusters \cite{haibo,kai,murayama}.

We remark that the SM couplings and non-linear interactions of the spin-2 particle can be important in determining the relic density of dark matter via thermal or non-thermal productions \cite{GMDM1,GMDM2,diphoton,cascade, GMetc,DD,GLDM} (See also Refs.~\cite{massless} for the case with a massless spin-2 mediator), and the non-linear interactions of the spin-2 particle are also crucial to extend the unitarity to a high scale \cite{dRGT,dRGT-review,adam,SB,sekhar}, in particular, in the case of a light spin-2 particle.
We also comment on the Vainshtein mechanism with non-linear interactions for screening the longitudinal degrees of freedom of a light spin-2 particle and point out its relevance for a large dark matter momentum in the Coulomb limit. 
In this work, we don't specify the non-linear interactions of the spin-2 particle at the quantitative level, but our discussion is valid in the Born and quantum regimes with a small dark matter momentum. Therefore, our results on the dark matter self-scattering with the effective Yukawa interactions is model-independent and insensitive to a concrete realization of non-linear interactions.

The paper is organized as follows. 
We first introduce the interactions of the massive spin-2 particle to dark matter of arbitrary spin. Then, we derive the self-scattering amplitudes for dark matter in the non-relativistic effective theory and identify the effective potentials for self-interactions. Next, we show the self-scattering cross sections for dark matter in the Born and non-perturbative limits and how the parameters of our model appear in the Sommerfeld factors.  The results for velocity-dependent self-scattering cross sections are presented for various choices of the parameters. There are four appendices dealing with the details on the self-scattering amplitudes, the non-relativistic operator basis for self-scattering, the useful momentum integral formulas for the effective potential as well as the details on the Sommerfeld factors. 
Finally, conclusions are drawn.

\section{Effective theory for dark matter self-interactions}

We provide the model setup for the interactions of the massive spin-2 mediator to dark matter and present the details on the self-scattering amplitudes for fermion, scalar and vector dark matter, in order.  As a result, we extend the Parity-invariant effective potentials for scalar and fermion dark matter in the literature with the case of the massive spin-2 mediator and obtain new results for vector dark matter.

\subsection{Setup for the massive spin-2 mediator}

We introduce the couplings of a massive spin-2 mediator ${\cal G}_{\mu\nu}$ to the SM particles and dark matter, through the energy-momentum tensor, as follows \cite{GMDM1,GMDM2,diphoton,cascade},
\bea
{\cal L}_{\rm int}= -\frac{c_{\rm SM}}{\Lambda} {\cal G}^{\mu\nu} T^{\rm SM}_{\mu\nu} -\frac{c_{\rm DM}}{\Lambda} {\cal G}^{\mu\nu} T^{\rm DM}_{\mu\nu}. \label{linear}
\eea
Here, $ T^{\rm SM}_{\mu\nu} , T^{\rm DM}_{\mu\nu}$ are the energy-momentum tensors for the SM particles and dark matter, respectively, $c_{\rm SM}, c_{\rm DM}$ are dimensionless couplings to the massive spin-2 mediator, and $\Lambda$ is the overall suppression scale.  

The mediator couplings for the SM particles are model-dependent.  For instance, they depend on the location of dark matter and the SM particles in the case of the warped extra dimension \cite{GMDM1,GMDM2}.
Based on the scattering amplitudes between dark matter of arbitrary spin and quarks, the systematic analysis of the DM-nucleon scattering cross section in the non-relativistic effective theory can be referred to our previous works in Ref.~\cite{DD,GLDM} and the results were applied to the direct detection with cosmic ray collisions \cite{DD2}.
As the SM couplings are not relevant for the DM self-scattering, we focus on the couplings of the massive spin-2 particle to dark matter in the later discussion.

Non-linear interactions for the massive spin-2 particle in massive gravity theories \cite{dRGT,massive,dRGT-review}, such as self-interactions with double gravitons, triple gravitons, etc, are important for a better UV behavior of the massive spin-2 particle, but they are model-dependent. Thus, we don't specify the non-linear interactions for the later discussion, and we just assume that the exchanges of the extra degrees of freedom appearing at the non-linear level could be suppressed sufficiently for the dark matter self-scattering. Thus, we focus on the linear interactions for the massive spin-2 mediator with physical degrees of freedom in a model-independent way.

We denote a real scalar dark matter, a Dirac fermion dark matter and a real vector dark matter by $S$, $\chi$ and $X$, respectively. Then, the energy-momentum tensor for each case of dark matter, $T^{\rm DM}_{\mu\nu}$, is given by
\bea
T^{ S}_{\mu\nu}&=&c_S\bigg[ \partial_\mu S \partial_\nu S-\frac{1}{2}g_{\mu\nu}\partial^\rho S \partial_\rho S+\frac{1}{2}g_{\mu\nu}  m^2_S S^2\bigg],\\
T^{\chi}_{\mu\nu}&=& c_\chi \bigg[\frac{i}{4}{\bar\chi}(\gamma_\mu\partial_\nu+\gamma_\nu\partial_\mu)\chi-\frac{i}{4} (\partial_\mu{\bar\chi}\gamma_\nu+\partial_\nu{\bar\chi}\gamma_\mu)\chi-g_{\mu\nu}(i {\bar\chi}\gamma^\mu\partial_\mu\chi- m_\chi {\bar\chi}\chi) \bigg]
\nonumber \\
&&+\frac{i}{2}g_{\mu\nu}\partial^\rho({\bar\chi}\gamma_\rho\chi)\bigg],  \\
T^{X}_{\mu\nu}&=&
c_X\bigg[ \frac{1}{4}g_{\mu\nu} X^{\lambda\rho} X_{\lambda\rho}+X_{\mu\lambda}X^\lambda\,_{\nu}+m^2_X\Big(X_{\mu} X_{\nu}-\frac{1}{2}g_{\mu\nu} X^\lambda  X_{\lambda}\Big)\bigg]
\eea
where $X_{\mu\nu}=\partial_\mu X_\nu-\partial_\nu X_\mu$ is the field strength tensor for vector dark matter.

The linear interactions for the massive spin-2 particle in eq.~(\ref{linear}) can be written in the momentum space, as follows,
\bea
\widetilde{{\cal L}}_{\rm int}(p) = -\frac{c_{\rm DM}}{\Lambda} \widetilde{{\cal G}}^{\mu\nu}(p) {T}^{\rm DM}_{\mu\nu}(-p). 
\eea
Then, the massive spin-2 particle mediates the self-scattering of dark matter, with the corresponding amplitude given in the following general form, 
\bea
{\cal M}=-\frac{c^2_{\rm DM}  }{\Lambda^2} \frac{i}{q^2-m^2_G}\,T^{\rm DM}_{\mu\nu}(q){\cal P}^{\mu\nu,\alpha\beta}(q) T^{\rm DM}_{\alpha\beta}(-q)  \label{ampl}
\eea
where $q$ is the 4-momentum transfer between dark matter particles and the tensor structure for the massive spin-2 propagator is given by
\bea
{\cal P}_{\mu\nu,\alpha\beta}(q)=\frac{1}{2}\Big(G_{\mu\alpha}G_{\nu\beta}+G_{\nu\alpha} G_{\mu\beta}- \frac{2}{3} G_{\mu\nu} G_{\alpha\beta}\Big)
\eea
with
\bea
G_{\mu\nu}\equiv \eta_{\mu\nu}- \frac{q_\mu q_\nu}{m^2_G}. 
\eea
Here, the energy-momentum tensor also depends on the momentum of dark matter, as will be shown explicitly in the next subsection.
We note that the sum of the spin-2 mediator polarizations is given by
\bea
\sum_s \epsilon_{\mu\nu}(q,s) \epsilon_{\alpha\beta}(q,s)= P_{\mu\nu,\alpha\beta}(q). 
\eea
The tensor $P_{\mu\nu,\alpha\beta}$ satisfies traceless and transverse conditions for on-shell spin-2 mediator, such as $\eta^{\alpha\beta} P_{\mu\nu,\alpha\beta}(q)=0$ and $q^\alpha P_{\mu\nu,\alpha\beta}(q)=0$.

Due to energy-momentum conservation, $q_\mu T^{\mu\nu}=0$, we can replace $G_{\mu\nu}$ in the scattering amplitude (\ref{ampl}) by $\eta_{\mu\nu}$. Then, the self-scattering amplitude in eq.~(\ref{ampl}) is divided into trace and traceless parts of energy-momentum tensor, as follows,
\bea
{\cal M}&=&-\frac{c^2_{\rm DM} }{2\Lambda^2} \frac{i}{q^2-m^2_G}\, \bigg(2 T^{\rm DM}_{\mu\nu} T^{{\rm DM},\mu\nu} -\frac{2}{3}(T^{\rm DM})^2 \bigg)  \nonumber \\
&=&-\frac{c^2_{\rm DM} }{2\Lambda^2} \frac{i}{q^2-m^2_G}\, \bigg(2{\tilde T}^{\rm DM}_{\mu\nu} {\tilde T}^{{\rm DM},\mu\nu} -\frac{1}{6}({T}^{\rm DM})^2\bigg) \label{ampl2}
\eea
where ${\tilde T}^{\rm DM}_{\mu\nu}$ is the traceless part of energy-momentum tensor given by ${\tilde T}^{\rm DM}_{\mu\nu}=T^{\rm DM}_{\mu\nu}-\frac{1}{4}\eta_{\mu\nu} {T}^{\rm DM}$ with ${T}^{\rm DM}$ being the trace of energy-momentum tensor.  

We note that only the $t$-channel amplitudes are taken into account in eq.~(\ref{ampl}) or (\ref{ampl2}), but they capture well the effective self-interactions for the self-scattering of non-identical dark matter particles and are valid for the forward self-scattering of dark matter. For identifying the effective fine-structure constants for dark matter self-interactions, it is enough to focus on the $t$-channel amplitudes. But, as will be discussed in Section 3, we also take into account the $u$-channel diagrams for identical particles and implement the symmetries of the scattering amplitudes.

\subsection{Fermion dark matter}

The energy-momentum tensor for a fermion DM $\chi$  is, in momentum space,
\bea
T^\chi_{\mu\nu}= -\frac{1}{4} {\bar u}_\chi(k_2)\Big(\gamma_\mu (k_{1\nu}+k_{2\nu})+\gamma_\nu (k_{1\mu}+k_{2\mu})-2\eta_{\mu\nu}(\slashed{k}_1+\slashed{k}_2-2m_\chi) \Big)u_\chi(k_1)
\eea
where the fermion DM is incoming into the vertex with momentum $k_1$ and is outgoing from the vertex with momentum $k_2$, and $u_\chi(k)$ is the wave function for fermion dark matter carrying the momentum $k$. 
Then, the trace of the energy-momentum tensor is given by
\bea
T^\chi=-\frac{1}{4} {\bar u}_\chi(k_2)\Big(-6(\slashed{k}_1+\slashed{k}_2)+16m_\chi\Big)u_\chi(k_1), \label{fdm-scalar}
\eea
whereas the traceless part of the energy-momentum tensor is given by
\bea
{\tilde T}^\chi_{\mu\nu}= -\frac{1}{4}{\bar u}_\chi(k_2)\Big(\gamma_\mu (k_{1\nu}+k_{2\nu})+\gamma_\nu (k_{1\mu}+k_{2\mu})-\frac{1}{2}\eta_{\mu\nu}(\slashed{k}_1+\slashed{k}_2) \Big) u_\chi(k_1). \label{fdm-tensor}
\eea

Therefore, using the results eq.~(\ref{feff}) for the self-scattering amplitude for fermion dark matter in Appendix A, we obtain  the effective self-interactions for fermion dark matter in terms of Galilean-covariant and Hermitian operators, $i{\vec q}$ and ${\vec v}^\perp$, and the spins of dark matter,  ${\vec s}_{1,2}$, as
\bea
\mathcal{L}_{\chi,\rm eff}&=& -\frac{1}{{\vec q}^2+m^2_G}\, \sum_{i} c^\chi_i {\cal O}_i \label{feff2}
\eea
with
\bea
{\cal O}_1&=&1, \\
{\cal O}_2&=&{\vec s}_1\cdot {\vec s}_2,  \\
{\cal O}_3&=&-\frac{1}{m^2_\chi}\,({\vec s}_1\cdot {\vec q})({\vec s}_2\cdot {\vec q}),  \\
{\cal O}_4 &=& ({\vec s}_1\cdot{\vec v}^\perp) ({\vec s}_2\cdot{\vec v}^\perp), \\
{\cal O}_5 &=&-\frac{i}{m_\chi}\,\bigg[({\vec s}_1\cdot {\vec q}) ({\vec s}_2\cdot{\vec v}^\perp)+ ({\vec s}_2\cdot {\vec q})({\vec s}_1\cdot{\vec v}^\perp) \bigg], \\
{\cal O}_7&=&-\frac{i}{m_\chi}\, ({\vec s}_1+{\vec s}_2)\cdot ({\vec q}\times {\vec v}^\perp)=-\frac{i}{m_\chi}\, ({\vec s}_1+{\vec s}_2)\cdot ({\vec q}\times {\vec v})
\eea
where
\bea
c^\chi_1&=& -\frac{c^2_\chi}{2 \Lambda^2}\, \Big(\frac{16}{3} m_{\chi}^4  + 8 m_{\chi}^2 {\vec q}^2 +8 m^4_\chi  (v^\perp)^2\Big), \\
c^\chi_2 &=&\frac{2c^2_\chi m^2_\chi}{\Lambda^2}\, {\vec q}^2, \\
c^\chi_3 &=& \frac{2c^2_\chi m^4_\chi}{\Lambda^2}, \\
c^\chi_4 &=& -\frac{2c^2_\chi m^2_\chi}{\Lambda^2}\, {\vec q}^2, \\
c^\chi_5 &=& \frac{2c^2_\chi m^3_\chi}{\Lambda^2}\, (i{\vec q}\cdot {\vec v}^\perp)=0, \\
c^\chi_7 &=&-\frac{4c^2_\chi m^4_\chi}{2\Lambda^2}.
\eea
The basis of non-relativistic operators for self-scattering \cite{tanedo} are listed in Appendix B for comparison.
The momentum transfer after scattering is given by ${\vec q}={\vec k}_1-{\vec k}_2={\vec p}_2-{\vec p}_1$ and the relative velocity between dark matter particles before scattering is given by $m_\chi {\vec v}={\vec p}_1-{\vec k}_1$, respectively. Then, the initial relative velocity ${\vec v}$ and the momentum transfer ${\vec q}$ are related by the transverse component of the relative velocity with respect to ${\vec q}$ as ${\vec v}^\bot={\vec v}+\frac{{\vec q}}{m_\chi}$, and we have the identities, ${\vec v}^\bot\cdot{\vec q}=0$ and ${\vec q}^2+m^2_\chi({\vec v}^\bot)^2=m^2_\chi v^2$.
We also note that a factor  $\int \frac{d^3p}{(2\pi)^3\sqrt{2E}} \, a^{(\dagger)}_\chi$ per each dark matter state, with dimension $E$, are to be multiplied as overall factors such that the above effective Lagrangian has a dimension 4. 

We note that the coefficient of ${\cal O}_5$ vanishes, and ${\cal O}_4$ is a velocity-suppressed operator with ${\vec q}^2$-suppressed coefficient, so it is unimportant for Sommerfeld enhancement. Therefore, we don't consider ${\cal O}_4$  in the effective potential below.

From the above result (\ref{feff2}) and the momentum integrals with the effective interactions listed in Appendix C, we get the effective potential for a pair of dark matter fermions \cite{tanedo},
\bea
V_{\chi,{\rm eff}}(r)&=&-\frac{1}{4m^2_\chi}\int \frac{d^3q }{(2\pi)^3}\,e^{i{\vec q}\cdot {\vec r}}\,  {\cal L}_{\chi,{\rm eff}}  \nonumber \\
&=& \frac{1}{4\pi r} \bigg\{ c_1(r) + c_2(r) ({\vec s}_1\cdot {\vec s}_2) + \frac{c_3(r)}{m^2_\chi r^2}\, \Big[3 ({\vec s}_1\cdot {\hat r})({\vec s}_2\cdot {\hat r})-{\vec s}_1\cdot {\vec s}_2 \Big] \nonumber \\
&&\quad+\frac{c_7(r)}{m_\chi r}\, ({\vec s}_1+ {\vec s}_2)\cdot ({\hat r}\times {\vec v})  \bigg\} \label{feff-full}
\eea 
where
\bea
c_1(r)&=& -\frac{2c^2_\chi m^2_\chi}{3 \Lambda^2}\, e^{-m_G r}\Big( 1+\frac{3}{2} (v^\perp)^2-\frac{3}{2}\Big(\frac{m_G}{m_\chi}\Big)^2 \Big), \label{fc1} \\ 
c_2(r)&=& -\frac{c^2_\chi m^2_G}{3\Lambda^2}\,e^{-m_G r} \Big(1-\frac{1}{2}(v^\perp)^2 \Big), \\
c_3(r)&=& \frac{c^2_\chi m^2_\chi}{2\Lambda^2} \,e^{-m_G r}\Big(1+m_G r+\frac{1}{3}(m_G r)^2 \Big)\Big(1+(v^\perp)^2 \Big), \\
c_7(r)&=& -\frac{c^2_\chi m^2_\chi}{\Lambda^2} \,e^{-m_G r} \Big(1-\frac{1}{8} \Big(\frac{m_G}{m_\chi}\Big)^2\Big)\Big(1+m_G r\Big).
\eea
Then, in terms of the total spin, ${\vec S}={\vec s}_1+{\vec s}_2$ and the orbital angular momentum, ${\vec L}={\vec r}\times {\vec p}$ with ${\vec p}=\frac{1}{2}m_\chi {\vec v}$, we can express the potential in the following \cite{tanedo},
\bea
V_{\chi,{\rm eff}}(r)=  \frac{1}{4\pi r} \bigg\{ c_1(r)-\frac{3}{4} c_2(r)+\frac{1}{2} c_2(r) {\vec S}^2 + \frac{c_3(r)}{2m^2_\chi r^2} \Big[3({\vec S}\cdot {\hat r})- {\vec S}^2\Big]+\frac{2c_7(r)}{m_\chi^2 r^2}\,{\vec S}\cdot {\vec L}  \bigg\}. \label{fermion-epot}
\eea

For non-relativistic fermion dark matter and $m_G\ll m_\chi$, we take the leading terms for the effective potential (\ref{fermion-epot}) to be proportional to $\frac{1}{r}$ as  
\bea
V_{\chi,{\rm eff}} &\simeq&  \frac{1}{4\pi r} \bigg( c_1(r)-\frac{3}{4} c_2(r) +\frac{1}{2}c_2(r) {\vec S}^2 \bigg)+\cdots \nonumber \\
&\simeq & - \frac{1}{4\pi r}\, e^{-m_G r}\, \cdot \frac{c^2_\chi m^2_\chi}{\Lambda^2} \bigg(\frac{2}{3}-\frac{5}{4}\Big( \frac{m_G}{m_\chi}\Big)^2  +\frac{1}{6} \Big( \frac{m_G}{m_\chi}\Big)^2 {\vec S}^2 \bigg) +\cdots.  \label{fc1}
\eea
Then, including the spin-dependent interactions for $ {\vec S}^2=s(s+1)$ with $s=1$ (triplet) and $s=0$ (singlet) states,  the effective potential  takes the following form,
\bea
V_{\chi,{\rm eff}} = - \frac{1}{4\pi r}\, e^{-m_G r}\,\left(\begin{array}{cc}  A_\chi+\frac{c^2_\chi m^2_\chi}{3\Lambda^2}\Big( \frac{m_G}{m_\chi}\Big)^2 & 0 \\ 0 & A_\chi \end{array} \right), \label{spin-F}
\eea
where the effective fine structure constant for fermion dark matter can be read from the $s=0$ interaction as follows,
\bea
A_\chi\simeq \frac{2c^2_\chi m^2_\chi}{3 \Lambda^2}\Big(1-\frac{15}{8} \Big( \frac{m_G}{m_\chi}\Big)^2  \Big). \label{fine-F}
\eea
As far as $m_G\ll m_\chi$, the effective potential is approximately independent of the spin states.

 We remark that the dipole interaction term, $c_3(r)$, and the spin-orbit interaction term, $c_7(r)$, were ignored in the approximation in eq.~(\ref{spin-F}), both of which have the $1/r^3$ dependence on the distance between dark matter particles. 
 Those $1/r^3$ terms look singular for $l=0$ channels as $r\rightarrow 0$, because the regularity condition, $r^2 \,V\rightarrow 0$, is not satisfied. Thus, after the regularization of the $1/r^3$ potential, the Sommerfeld factor for the dipole interaction term  can be large at the resonance when the pair of dark matter particles form a bound state \cite{tanedo,pseudoscalar}. However,  we can treat such $1/r^3$ terms perturbatively as the effective Yukawa interaction is dominant in the expansion of the potential with $1/r^n$,  just as the fine splitting due to the spin-orbit coupling can be dealt with as perturbation in hydrogen atom \cite{kai,reece}. Thus, we take the $1/r^3$ terms to be sub-dominant for Sommerfeld effects.
Moreover, in the effective theory for a massive spin-2 mediator, the $1/r^3$ terms have the coefficient given by $3({\vec s}_1\cdot{\hat r})({\vec s}_2\cdot {\hat r})-{\vec s}_1\cdot {\vec s}_2$, thus they vanish upon the action of the operator on $l=0$ channels for $s=0$ singlet state \cite{reece}. The dipole operator mixes $l=j\pm 1$ states, resulting in the mixing between $l=0$ and $l=2$ channels for $s=1$ triplet state \cite{tanedo,pseudoscalar,reece}, but the $l=0$ and $l=2$ channels do not give rise to strong enhancements in the perturbative regime \cite{reece}. But, it will be interesting to study the effects of the level mixing effects in the non-perturbative regime.  Similarly unsuppressed dipole-dipole interactions for vector dark matter exist and appropriate comments will be made in the next section. 
 
We also note that the spin-orbit coupling $c_7(r)$ is ignored in the approximation in eq.~(\ref{fc1}), because it is suppressed doubly by $1/r^3$ and the velocity of dark matter. 
Therefore, it is enough to focus on the effective Yukawa potential for the later discussion on the dark matter self-scattering.

\subsection{Scalar dark matter}

The energy-momentum tensor for  a scalar DM $S$ is, in momentum space,
\bea
T^S_{\mu\nu}= - \Big( m^2_S \eta_{\mu\nu}+C_{\mu\nu,\alpha\beta} k^\alpha_1 k^\beta_2 \Big) \label{scalar-emtensor}
\eea
where 
\bea
C_{\mu\nu,\alpha\beta}\equiv \eta_{\mu\alpha}\eta_{\nu\beta}+\eta_{\nu\alpha} \eta_{\mu\beta}-\eta_{\mu\nu}\eta_{\alpha\beta}
\eea
and the scalar DM is incoming into the vertex with momentum $k_1$ and is outgoing from the vertex with momentum $k_2$. 
Then, the trace of the energy-momentum tensor is given by
\bea
T^S=-\Big(4m^2_S - 2 (k_1\cdot k_2)\Big), \label{scalar-trace}
\eea
and the traceless part of the energy-momentum tensor is given by
\bea
{\tilde T}^S_{\mu\nu}= -\Big(k_{1\mu}k_{2\nu}+k_{2\mu}k_{1\nu}-\frac{1}{2}\eta_{\mu\nu}(k_1\cdot k_2) \Big).   \label{sdm-tensor}
\eea

Then, the effective self-interaction Lagrangian for scalar dark matter are given by
\bea
\mathcal{L}_{S,\rm eff}=-\frac{1}{{\vec q}^2+m^2_G}\,  c^S_1 {\cal O}_1 \label{seff2}
\eea
where
\bea
c^S_1=-\frac{c^2_S}{2\Lambda^2}\Big( \frac{16}{3}m_S^4 +\frac{8}{3}m_S^2 \vec{q}^2+8m_S^4 (v^\perp)^2\Big). 
\eea
As a result,  from the above result (\ref{seff2}) and the momentum integrals with the effective interactions listed in Appendix C, we get the effective potential for a pair of dark matter scalars, 
\bea
V_{S,{\rm eff}} \simeq -\frac{A_S}{4\pi r}\, e^{-m_G r}
\label{sc1}
\eea
where the effective fine structure constant for scalar dark matter is given by
\bea
A_S\simeq \frac{2c^2_S m^2_S}{3 \Lambda^2}\Big(1-\frac{1}{2}\Big( \frac{m_G}{m_S}\Big)^2  \Big). \label{fine-S}
\eea
Thus, the Yukawa-type potential for scalar dark matter in eq.~(\ref{sc1}) has the same power dependence in the mass of the spin-2 particle, as compared to the case for fermion dark matter in eq.~(\ref{fc1}), but with a different coefficient.

\subsection{Vector dark matter}

The energy-momentum tensor for a vector DM $X$  is, in momentum space,
\bea
T^X_{\mu\nu}=-\Big(m^2_X C_{\mu\nu,\alpha\beta}+W_{\mu\nu,\alpha\beta} \Big) \epsilon^{\alpha}(k_1)\epsilon^{*\beta}(k_2)
\eea
where $\epsilon^\alpha(k)$ is the polarization vector for the vector DM and 
\bea
W_{\mu\nu,\alpha\beta}&\equiv & -\eta_{\alpha\beta} k_{1\mu} k_{2\nu} -\eta_{\mu\alpha} (k_1\cdot k_2\,\eta_{\nu\beta}-k_{1\beta} k_{2\nu})+\eta_{\mu\beta} k_{1\nu} k_{2\alpha} \nonumber \\
&& -\frac{1}{2}\eta_{\mu\nu}(k_{1\beta}k_{2\alpha}-k_1\cdot k_2\, \eta_{\alpha\beta}) +(\mu\leftrightarrow\nu ).
\eea
Likewise as before, the vector DM  is incoming into the vertex with momentum $k_1$ and is outgoing from the vertex with momentum $k_2$.
Then, the trace of the energy-momentum tensor is given by
\bea
T^X=2 m^2_X \eta_{\alpha\beta} \epsilon^{\alpha}(k_1)\epsilon^{*\beta}(k_2).  \label{vdm-scalar}
\eea
\bea
{\tilde T}^X_{\mu\nu}=-\Big(m^2_X C_{\mu\nu,\alpha\beta}+W_{\mu\nu,\alpha\beta}+\frac{1}{2}m^2_X \eta_{\mu\nu}\eta_{\alpha\beta} \Big) \epsilon^{\alpha}(k_1)\epsilon^{*\beta}(k_2).\label{vdm-tensor}
\eea
Here, we note that $W_{\mu\nu,\alpha\beta} \eta^{\mu\nu}=0$, due to the fact that the energy-momentum tensor for transverse polarizations of vector dark matter is trace-free.

To compute the effective potential for vector dark matter, we take the on-shell polarization vector for vector dark matter with momentum $p^\mu=(E_p,{\vec p})$,
\bea
\epsilon^{s\mu}(p)=
\begin{pmatrix}
  {1\over m_X}\, {\vec p}\cdot \vec{e}_s \\ 
  \Big(1-(1-\frac{E_{p}}{m_X})\delta_{3s}\Big)\vec{e}_s
\end{pmatrix}
\simeq \begin{pmatrix}
  {1\over m_X}\, {\vec p}\cdot \vec{e}_s \\ 
  \vec{e}_s
\end{pmatrix},
\quad s=1,2,3
\eea
where $s=3$ corresponds to the longitudinal polarization and $s=1,2$ stand for the transverse polarizations.
Then, the polarization vectors for incoming and outgoing vector dark matter particles are given by
\begin{equation} 
\epsilon^{s\mu}(k_1)\simeq
\begin{pmatrix}
  {1\over 2m_X}(\vec{K}+\vec{q})\cdot \vec{e}_s \\ 
\vec{e}_s
\end{pmatrix},\ \ 
\epsilon^{s'\mu}(k_2)\simeq
\begin{pmatrix}
  {1\over 2m_X}(\vec{K}-\vec{q})\cdot \vec{e'}_{s'} \\ 
 \vec{e'}_{s'}
\end{pmatrix}
\end{equation}
\begin{equation} 
\epsilon^{r\mu}(p_1)\simeq
\begin{pmatrix}
  {1\over 2m_X}(\vec{P}-\vec{q})\cdot \vec{e}_r \\ 
\vec{e}_r
\end{pmatrix},\ \ 
\epsilon^{r'\mu}(p_2)\simeq
\begin{pmatrix}
  {1\over 2m_X}(\vec{P}+\vec{q})\cdot \vec{e'}_{r'} \\ 
 \vec{e'}_{r'}
\end{pmatrix}
\end{equation}
where we have used ${\vec K}={\vec k}_1+{\vec k}_2$, ${\vec P}={\vec p}_1+{\vec p}_2$, and ${\vec q}={\vec k}_1-{\vec k}_2$, and the normalization conditions for polarization 3-vectors are ${\vec e}_r\cdot {\vec e}_{r'}=\delta_{rr'}$ and $\vec{e}_{s}\cdot \vec{e'}_{s'}=\delta_{s s'}$.
Then, we can construct the spin operators for vector dark matter \cite{V-DD} as
 \bea
 \vec{S}_X^{s's}&=&i \vec{e}_{s}\times \vec{e'}_{s'} \\
 {\cal S}_{ij}^{s's}&=& {1\over 2}(e_{si}e'_{s'j}+e_{sj}e'_{s'i}),
 \eea
 and $\vec{q}\cdot {\cal S}^{r'r}\cdot \vec{q}=q_i \, {\cal S}_{ij}^{r'r} \, q_j$, etc, 

As a consequence, from the results in eq.~(\ref{VDM-M}) in Appendix A and the momentum integrals with the effective interactions listed in Appendix C, the effective self-interaction Lagrangian for vector dark matter leads to
the effective potential for a pair of vector dark matter,
\bea
V_{X,{\rm eff}}(r)
&=&{1 \over 4\pi r}\Big\{ c_1(r)\,\delta^{s's}\delta^{r'r} +c_2(r)\, (\vec{S}_X^{s's}\cdot \vec{S}_X^{r'r}) \nonumber \\
&&\quad+\frac{1}{2} c_2(r) \Big(({\vec v}^\perp\cdot {\cal S}^{s's}\cdot {\vec v}^\perp)\, \delta^{r'r} +({\vec v}^\perp\cdot {\cal S}^{r'r}\cdot {\vec v}^\perp)\, \delta^{s's}  \Big) \nonumber  \\
&&\quad +\frac{c_3(r)}{m^2_X r^2} \Big[  \frac{1}{3}\Big(3  {\hat r} \cdot(\delta^{s's}  {\cal S}^{r'r}+ \delta^{r'r} {\cal S}^{s's} )\cdot {\hat r} -2\delta^{s's} \delta^{r'r}\Big) \label{Xpot}\\
&&\qquad \quad
+3  {\hat r} \cdot ( {\cal S}^{r's}_{ii}{\cal S}^{rs'}+{\cal S}^{rs'}_{ii} {\cal S}^{r's}-{\cal S}^{rs}_{ii}{\cal S}^{r's'}- {\cal S}^{r's'}_{ii}{\cal S}^{rs}) \cdot {\hat r} -2  {\vec S}^{s's}_X\cdot {\vec S}^{r'r}_X \Big]  \nonumber  \\
&&\quad +\frac{c_7(r)}{m_X r}(\vec{S}_X^{ss'}\delta^{s's}+\vec{S}_X^{rr'}\delta^{s's})\cdot(\hat{r}\times \vec{v}) \Big\}.  \nonumber 
\eea
where
\bea
c_1(r)&=& -{2 c^2_Xm_X^2 \over 3 \Lambda^2 } \,e^{-m_G r} \Big(1+\frac{3}{2}(v^\perp)^2+\frac{1}{6}\Big( \frac{m_G}{m_X}\Big)^2 \Big), \\
c_2(r)&=&-\frac{c^2_X m^2_G}{3\Lambda^2}\, e^{-m_G r}, \\
c_3(r) &=&-\frac{c^2_Xm^2_X}{2\Lambda^2} \,e^{-m_G r}\Big(1+m_G r +\frac{1}{3} (m_G r)^2 \Big), \\
c_7(r)&=& \frac{5c^2_X m^2_X}{6\Lambda^2}\, e^{-m_G r} (1+m_G r).
\eea
Therefore, the effective potential has a similar form as in the case for fermion dark matter in eq.~(\ref{feff-full}), but there are more structures in  dipole-dipole interactions due to new effective spin-dependent interactions for vector dark matter, as in the recent discussion on the direct detection scattering cross section in the effective theory for vector dark matter  \cite{V-DD}.
We find that the general correlations between spins of a pair of vector dark matter appear in the effective potential (\ref{Xpot}), although the corresponding potentials are suppressed by the extra power with $1/r^2$ as compared to the Yukawa potential.

For a light spin-2 mediator with $m_G\ll m_X$ and a small velocity of dark matter, the leading form of the effective potential is given by 
\bea
V_{X,{\rm eff}} &\simeq &\frac{1}{4\pi r} \bigg(c_1(r)-2c_2(r) + \frac{1}{2}\, c_2(r) {\vec S}^2_X+\cdots \bigg) \nonumber \\
& \simeq &-\frac{1}{4\pi r}\, e^{-m_G r} \,\cdot \frac{c^2_X m^2_X}{\Lambda^2} \bigg(\frac{2}{3}-\frac{5}{9} \Big(\frac{m_G}{m_X} \Big)^2+ \frac{1}{6}\Big(\frac{m_G}{m_X} \Big)^2 {\vec S}^2_X\bigg)+\cdots. \label{Xeff}
\eea
Therefore, due to the spin-dependent interactions  for $ {\vec S}^2_X=s(s+1)$ with $s=2$ (quintuplet), $s=1$ (triplet), and $s=0$ (singlet) states, the effective Yukawa potential for vector dark matter takes the following form, 
\bea
V_{X,{\rm eff}} \simeq - \frac{1}{4\pi r}\, e^{-m_G r}\,\left(\begin{array}{ccc}  A_X+\frac{c^2_X m^2_X}{\Lambda^2}\Big( \frac{m_G}{m_X}\Big)^2 & 0 & 0 \\  0 & A_X +\frac{c^2_X m^2_X}{3\Lambda^2}\Big( \frac{m_G}{m_X}\Big)^2& 0\\ 0 &  0 & A_X \end{array} \right), \label{spin-V}
\eea
where the $s=0$ interaction leads to the effective fine structure constant for vector dark matter,  given by
\bea
A_X\simeq  \frac{2c^2_X m^2_X}{3\Lambda^2} \bigg(1-\frac{5}{6} \Big(\frac{m_G}{m_X} \Big)^2\bigg). \label{fine-V}
\eea
Here, we have taken the total spin operator for vector dark matter by ${\vec S}_X={\vec S}_{X,1}+{\vec S}_{X,2}$.
We note that the spin wave functions for $s=2, s=0$ states are symmetric and the one for $s=1$ is antisymmetric. 
As far as $m_G\ll m_\chi$, the above effective potential is approximately independent of the spin states.

As compared to the spin-dependent interactions for fermion dark matter in eqs.~(\ref{fc1}) or (\ref{spin-F}), the ${\vec S}^2_{\rm DM}$ terms appear with the same coefficient but the eigenvalues of the effective potential for the spin non-singlet states differ from the one for fermion dark matter. Therefore, the resulting effective Yukawa potential for vector dark matter has different $(m_G/m_{\rm DM})^2$ corrections for both spin singlet and non-singlet states.

We also remark that the dipole interactions with $c_3(r)$ coefficient and $1/r^3$ dependence were ignored in eq.~(\ref{spin-V}) in the perturbative expansion of the potential with higher order terms of $1/r^n$, as in the case for the similar dipole interactions for fermion dark matter discussed in the previous subsection.

\subsection{Discussion on the effective potentials for dark matter}

The effective fine structure constant for vector dark matter in eq.~(\ref{fine-V}) depends on the mass of the spin-2 mediator differently, as compared to those for fermion or scalar dark matter, given in eqs.~(\ref{fine-F}) and (\ref{fine-S}), respectively. But, the spin-dependent Yukawa potential are of the same form in the squared total spin operator for fermion and vector dark matter as shown in eqs.~(\ref{fc1}) and (\ref{Xeff}) up to the $(m_G/m_{\rm DM})^2$ terms. 
For $m_G\ll m_{\rm DM}$, the spin-dependent terms can be ignored and the effective fine structure constant in the effective Yukawa potential depends only on the DM mass, which was the focus of the work in Ref.~\cite{SB}.

 \begin{table}[h]
\begin{center}
\begin{tabular}{c|c|c|c|c}
\hline
 {\rm mediator} & {\rm interaction}   & $1/r$  & $ ({\vec s}_1\cdot {\vec s}_2)/r$  & $ D_{12}/r^3 $ \\ 
\hline
\hline 
{\rm scalar} &  $ \lambda_s {\bar\chi}\chi \,s$  & $-\lambda^2_s$ &  0  &  0 \\ 

{\rm pseudoscalar} &  $i\lambda_a {\bar\chi}\gamma^5 \chi a$  &  0 &  $\frac{\lambda^2_a m^2_a}{3m^2_\chi}$  & $\frac{\lambda^2_a}{m^2_\chi} h(m_a,r)$ \\ 

{\rm Goldstone} & $\frac{1}{f}{\bar\chi}\gamma^\mu \gamma^5\chi \partial_\mu a$   & 0   & $\frac{4 m^2_a}{3f^2}$  & $\frac{4}{f^2} h(m_a,r)$  \\ 

{\rm vector} &  $g_v {\bar\chi}\gamma^\mu\chi A_\mu$  & $\pm g^2_v \Big(1+\frac{m^2_A}{4m^2_\chi}\Big)$  & $\pm \frac{2g^2_v m^2_A}{3m^2_\chi}$   & $\mp \frac{g^2_v}{m^2_\chi} h(m_A,r)$ \\ 

{\rm axial\, vector} & $g_a {\bar\chi}\gamma^\mu\gamma^5 \chi A_\mu$   & 0 & $-\frac{8 g^2_a}{3}\Big(1-\frac{m^2_A}{8m^2_\chi} \Big)$  & $g^2_a \Big(\frac{1}{m^2_\chi}+\frac{4}{m^2_A} \Big)h(m_A,r)$ \\ 

{\rm field\, strength} & $\frac{i}{2\Lambda} {\bar\chi}\sigma^{\mu\nu}\chi F_{\mu\nu}$   & 0  &  $\mp \frac{2m^2_A}{3\Lambda^2}$  &  $\pm \frac{1}{\Lambda^2} h(m_A,r)$ \\
\hline
{\rm graviton} &$-\frac{1}{\Lambda}T^{\chi,\mu\nu}G_{\mu\nu}$  & $-\frac{2m^2_\chi}{3 \Lambda^2} \Big(1-\frac{3}{2}\frac{m_G^2}{m_S^2}\Big) $   &    $-\frac{m^2_G}{3\Lambda^2} $ & $\frac{1}{2\Lambda^2} h(m_G, r)$ \\ 
{\rm graviton} &$-\frac{1}{\Lambda}T^{S,\mu\nu}G_{\mu\nu}$  & $-\frac{2 m^2_S}{3 \Lambda^2} \Big(1-\frac{1}{2}\frac{m_G^2}{m_S^2}\Big)$   &    0 & 0 \\ 
{\rm graviton} &$-\frac{1}{\Lambda}T^{X,\mu\nu}G_{\mu\nu}$  &  $-\frac{2 m^2_X}{3 \Lambda^2} \Big(1+\frac{1}{6}\frac{m_G^2}{m_X^2}\Big)$   &    
$-\frac{m^2_G}{3\Lambda^2}$    &  $-\frac{1}{2\Lambda^2}\, h(m_G, r)$  \\ 
\hline
\end{tabular}
\caption{Parity-invariant effective potential for fermion dark matter with scalar or vector mediators \cite{tanedo} as well as dark matter of arbitrary spin with the massive spin-2 mediator obtained newly in this work. Each term contains implicitly a Yukawa factor, $e^{-mr}/(4\pi)$, with $m=m_s, m_a, m_A, m_G$, and $h(m,r)=1+m r+\frac{1}{3}(m r)^2$, and the magnetic dipole-dipole operator for fermion dark matter is $D_{12}\equiv 3({\vec s}_1\cdot{\hat r})({\vec s}_2\cdot {\hat r})-{\vec s}_1\cdot {\vec s}_2$.  In the case of vector dark matter, the spin-dependent potential is described by new operators as given in eq.~(\ref{Xpot}) and explained in the text. } 
\label{table:effpot}
\end{center}
\end{table}

In Table~\ref{table:effpot}, we summarize our results for the effective potential for dark matter with various mediators, including scalar and vector mediators known in the literature \cite{tanedo} as well as the massive spin-2 mediator in our case. In the case with spin-2 mediator, all the velocity-independent interactions are nonzero for fermion and vector dark matter, but only the spin-independent potential  survives for scalar dark matter. For $m_G\ll m_{\rm DM}$ in the Coulomb limit, the spin-independent effective potentials are approximately the same, independent of the spins of dark matter, so the non-perturbative and Sommerfeld effects become similar.
On the other hand, we identify the spin-dependent interactions for fermion and vector dark matter: spin-spin interactions are in common but dipole-dipole interactions for vector dark matter contain new effective operators with general correlations of spins. Here, we note that the $1/r$ potentials  for fermion and vector dark matter in Table~\ref{table:effpot} are read off from $c_1(r)$ in eqs.~(\ref{feff-full}) and (\ref{Xeff}), respectively.

\section{Velocity-dependent self-scattering}

In this section, we discuss the Born limit of the self-scattering cross section and comment on the classical regime.
Then, we take into account the Sommerfeld effects by considering the approximate analytic solution with  the Hulth\'en potential. The parameter space for the self-scattering cross section to solve the small-scale problems at galaxies is also commented upon.

First, we note that the momentum transfer cross section for DM self-scattering \cite{haibo,kai} is given by
\bea
\sigma_T = 2\pi \int^1_{-1}\frac{d\sigma}{d\Omega}\, \Big(1-|\cos\theta|\Big) d\cos\theta. \label{ssx}
\eea
When there is a $t$-channel only, we can replace $1-|\cos\theta|$ just by $1-\cos\theta$. But, when there are both $t$-channel and $u$-channel such as for the scattering of identical states, we need to consider eq.~(\ref{ssx}).

The differential scattering cross section in the partial-wave expansions can be written as
\bea
\frac{d\sigma}{d\Omega} =|f(k,\theta)|^2  \label{diffcross}
\eea
with
\bea
f(k,\theta) =\sum_{l=0}^\infty (2l+1) f_l(k) P_l(\cos\theta)
\eea
where the partial-wave amplitude $f_l(k)$ is given in term of the phase shift, $\delta_l(k)$,
\bea
f_l(k) =\frac{e^{2i\delta_l(k)}-1}{2ik} =\frac{1}{k(\cot\delta_l(k)-i)},
\eea
and $k$ is the momentum of dark matter in the center of mass frame.
In the case for identical particles, we only have to replace $f(k,\theta)$ by $f(k,\theta)+\xi f(k,\pi-\theta)$ with $\xi=+1$ for space-symmetric wave function and $\xi=-1$ for space-antisymmetric wave function.
For the self-scattering for spin-$\frac{1}{2}$ fermion dark matter, we need to take into account the spin-states for the system, namely, the statistical distribution $\frac{1}{4}$ for spin singlet with space-symmetric wave function and $\frac{3}{4}$ for spin triplet with space-antisymmetric wave function  \cite{kai}. For the self-scattering of identical vector dark matter particles,  similar symmetric properties of the space wave functions should be take into account, that is, the $s=0, 2$ states are space-symmetric and the $s=1$ states are space-antisymmetric.

As a consequence, the differential scattering cross sections for identical particles for scalar, fermion and vector dark matter are given, respectively, by
\bea
\frac{d\sigma_{SS}}{d\Omega}&=& \Big|f^S(k,\theta)+  f^S(k,\pi-\theta)\Big|^2, \\
\frac{d\sigma_{\chi\chi}}{d\Omega}&=& \frac{1}{4} \bigg(\Big|f^\chi_{s=0}(k,\theta)+  f^\chi_{s=0}(k,\pi-\theta)\Big|^2+3\Big|f^\chi_{s=1}(k,\theta)-  f^\chi_{s=1}(k,\pi-\theta)\Big|^2  \bigg), \\
\frac{d\sigma_{XX}}{d\Omega}&=& \frac{1}{9} \bigg(\Big| f^X_{s=0}(k,\theta)+  f^X_{s=0}(k,\pi-\theta)\Big|^2  \nonumber \\
&& +3\Big|f^X_{s=1}(k,\theta)-  f^X_{s=1}(k,\pi-\theta)\Big|^2+5\Big| f^X_{s=2}(k,\theta)+  f^X_{s=2}(k,\pi-\theta)\Big|^2 \bigg).
\eea
On the other hand, for the self-scattering between particle and anti-particle, such as $\chi{\bar\chi}\to \chi{\bar\chi}$, we can use 
\bea
\frac{d\sigma_{\chi{\bar\chi}}} {d\Omega}=\frac{1}{4} \bigg( \Big|f^\chi_{s=0}(k,\theta)\Big|^2+3\Big |f^\chi_{s=1}(k,\theta)\Big|^2 \bigg).
\eea
Here, $f^S(k,\theta), f^\chi_s(k,\theta), f^X_s(k,\theta) $ with $s$ being the total spin of dark matter particles are the (spin-dependent) scattering amplitudes without wave-function symmetries taken into account.

\subsection{Born regime}

For a Yukawa-type potential, $V=-\frac{A_{\rm DM}}{4\pi r} \, e^{-m_Gr}$, we first consider the perturbative regime with weak self-interaction for dark matter, $A_{\rm DM} m_{\rm DM}/(4\pi m_G)\lesssim 1$. This is the Born regime. 

We note that the Bohr radius of dark matter bound states is given by $r_B=1/(\alpha_{\rm DM} m_{\rm DM})$ with $\alpha_{\rm DM}=A_{\rm DM}/(4\pi)$ and the range of the Yukawa potential is given by $r_G=m^{-1}_G$. Thus, we define the regime with $r_B>r_G$ to be the perturbative regime as there would be no bound states in this case. On the other hand, the regime with $r_B<r_G$ is called the non-perturbative regime where dark matter bound states are possible. Although the presence of bound states does not necessarily imply the breakdown of perturbativity as in usual atoms, we often need to go beyond the Born  regime in order to describe the bound-state formation and/or the non-perturbative effects with a massive mediator particle, which will be the regime that we will consider in the next subsection. 

The differential cross section for ${\rm DM\, DM\rightarrow DM\,DM}$, the elastic self-scattering of non-identical states, with $t$-channel dominance, is independent of the spins of dark matter, given by
\bea
\frac{d\sigma_{{\rm self}}}{d\Omega} = \frac{A^2_{\rm DM} m^2_{\rm DM}}{16\pi^2}\, \frac{1}{(m^2_G+m^2_{\rm DM} v^2(1-\cos\theta)/2)^2}. \label{tchannel}
\eea
Thus, inserting eq.~(\ref{tchannel}) into eq.~(\ref{ssx}), the corresponding momentum transfer cross section for the DM self-scattering becomes in the Born regime
\bea
\sigma^{\rm Born}_{T}= \frac{A^2_{\rm DM} }{2\pi m^2_{\rm DM} v^4}\, \left[ \ln\Big(1+\frac{m^2_{\rm DM} v^2}{m^2_G} \Big)-\frac{m^2_{\rm DM} v^2}{m^2_G+m^2_{\rm DM} v^2} \right]. \label{born}
\eea
Then, for $m_{\rm DM} v\lesssim m_G$, we can ignore the momentum transfer for the DM self-scattering, so the velocity expansion of the above result (\ref{born})  leads to $\sigma^{\rm Born}_{T}\approx\frac{A^2_{\rm DM} m^2_{\rm DM}}{4\pi m^4_G} \approx \frac{c^4_{\rm DM} m^6_{\rm DM}}{9\pi \Lambda^4 m^4_G}$ where the latter approximation is valid for $m_G\ll m_{\rm DM}$.

But, when there are scattering processes for identical dark matter particles, it is important to keep the $u$- and/or $s$-channels as well.
Indeed, the self-scattering cross sections with $t$-channels only should be replaced by the full DM self-scattering cross sections, so the resulting momentum transfer cross sections depend on the spins of dark matter.
Taking the limit of a small DM velocity with $m_{\rm DM} v\lesssim m_G$ in the scattering amplitudes apart from the spin-2 mediator propagators, we obtain the Born cross sections \cite{SB} for fermion, scalar and vector dark matter in order as follows,
\bea
\sigma_{S, T}^{\rm Born}&\simeq &{c_{S}^4m_{S}^2\over 9\pi \Lambda^4 r_S v^2 }{\ln\Big(1+{m_S^2 v^2 \over m_G^2}\Big)\over \Big(1 + {m_S^2v^2\over 2m_G^2}\Big)^3},  \label{born-S} \\
\sigma^{\rm Born}_{\chi,T}&=&{1\over 4}\Big(2\sigma^{\rm Born}_{\chi\chi,T }+\sigma^{\rm Born}_{\chi\bar{\chi}.T}\Big) \nonumber \\
&\simeq&{c_\chi^4 m_\chi^2\ \over 18\pi \Lambda^4 v^4}\left[ \bigg(1+ {2m_\chi^2 m_G^4 v^2 \over(m_\chi^2 v^2 + 2m_G^2)^3} \bigg) \ln\Big(1+{m_\chi^2 v^2 \over m_G^2} \Big) -{m_\chi^2 v^2 \over m_\chi^2 v^2 +m_G^2} \right],  \label{born-F} \\
\sigma_{X,T}^{\rm Born}&\simeq&{c_X^4 m_{X}^2 \over 27\pi \Lambda^4 r_X v^2}{(32-56r_X+27r_X^2)\over (4-r_X)^2}\,{\ln\Big(1+{m_X^2 v^2 \over m_G^2}\Big)\over \Big(1 + {m_X^2v^2\over 2m_G^2}\Big)^3}  \label{born-V}
\eea
with
\bea
\sigma_{\chi\chi,T}^{\rm Born}&\simeq&{c_{\chi}^4m_{\chi}^2\over 36\pi \Lambda^4 r_\chi v^2 }{\ln\Big(1+{m_\chi^2 v^2 \over m_G^2}\Big)\over \Big(1 + {m_\chi^2v^2\over 2m_G^2}\Big)^3},  \\
\sigma_{{\chi\bar{\chi}},T}^{\rm Born}&\simeq & {2c_{\chi}^4m_{\chi}^2\over 9\pi \Lambda^4 v^4}\bigg[ \ln\Big(1+{m_{\chi}^2 v^2\over m_G^2} \Big)-{m_{\chi}^2 v^2\over m_G^2+m_{\chi}^2 v^2}  \bigg].
\eea
Here, $r_{\rm DM}\equiv (m_G/m_{\rm DM})^2$ with ${\rm DM}=S, \chi, X$.
We note that the $s$-channel contributions to the self-scattering of scalar or fermion dark matter are velocity-suppressed by the overall factor, so it is ignored in the Born limit. On the other hand, the $s$-channel contribution to the counterpart of vector dark matter is not velocity-suppressed, so it is included in the above results.

As a result, in the Born regime with a vanishing DM velocity, the momentum transfer self-scattering cross sections divided by the DM mass  are further approximated to
\bea
\frac{\sigma_{S,T}^{\rm Born}}{m_S}&\simeq& {c_S^4 m_S\over 9\pi\Lambda^4 r_S^2},  \\
\frac{\sigma_{\chi,T}^{\rm Born}}{m_\chi}&\simeq & {c_\chi^4 m_\chi \over 24\pi \Lambda^4 r_\chi^2}, \\
\frac{\sigma_{X,T}^{\rm Born}}{m_X}&\simeq& {c_X^4 m_X \over 27 \pi \Lambda^4}{(32-56r_X+27r_X^2) \over r_X^2 (4-r_X)^2}.
\eea
We note that the overall factors in the above results differ from the total self-scattering cross sections \cite{GLDM}, by $1/2, 3/4$ and $1/2$, for scalar, fermion and dark matter cases, respectively. These are due to the inclusion of the forward scattering with equal weight for the total self-scattering cross sections, which would lead to no effect on the dark matter distribution.

\subsection{Non-perturbative effects on self-scattering}

We now consider the non-perturbative regime with $A_{\rm DM} m_{\rm DM}/(4\pi m_G)\gtrsim 1$, but $m_{\rm DM} v\lesssim  m_G$. This is the quantum regime.
In this case, the quantum mechanical effects such as bound states of dark matter can be important. Moreover, even if there is no bound state, the self-scattering cross section for dark matter can be enhanced by the resonance effects.
We also discuss the self-scattering in the effective range theory for a non-relativistic dark matter with a focus on the Hulth\'en potential approximation and match the results to the exact Born cross sections.

For the Yukawa-type potential, $V=-\frac{A_{\rm DM}}{4\pi r} \, e^{-m_Gr}$, there is no analytic solution for the Sommerfeld factor. But, we can adopt the approximate analytic solutions \cite{cassel} by replacing it with the Hulth\'en potential,
\bea
V_H= -\frac{A_{\rm DM}}{4\pi} \frac{\delta e^{-\delta r}}{1-e^{-\delta r}} \label{hulthen}
\eea
where the parameter $\delta$ is matched by $\delta=\frac{\pi^2}{6} m_G$ from the condition that the first moment of the potential, $\int^\infty_0 r' V(r') dr$, is unchanged for $V\rightarrow V_H$. Then, for $\delta r\ll 1$, we have $V_H\approx -\frac{A_{\rm DM}}{4\pi r}$.
For a nonzero angular momentum of the system with a pair of DM particles, i.e. $l\neq 0$, we also replace the centrifugal term $V_l=\frac{l(l+1)}{m_{\rm DM} r^2}$ by 
\be
{\tilde V}_l=\frac{l(l+1)}{m_{\rm DM}} \frac{\delta^2 e^{-\delta r}}{(1-e^{-\delta r})^2}. 
\ee
For $\delta r\ll 1$, similarly we get ${\tilde V}_l\approx V_l$. 

Adopting the Hulth\'en potential as the approximation for the Yukawa potential,  for the $s$-wave dominance, we obtain the exact result for the total self-scattering cross section \cite{haibo2}, as follows,
\bea
\sigma^{\rm Hulthen}_{\rm self}\simeq \frac{4\pi \sin^2\delta_0}{k^2}. \label{selfH}
\eea 
where the phase shift for the $s$-wave is given by
\bea
\delta_0={\rm arg} \left(\frac{i\Gamma(\lambda_++\lambda_--2)}{\Gamma(\lambda_+)\Gamma(\lambda_-)} \right) \label{phaseshift}
\eea
with
\bea
\lambda_\pm=1 + \frac{ik}{\delta} \pm \sqrt{\eta^2-\frac{k^2}{\delta^2}}, \qquad \eta=\sqrt{ \frac{A_{\rm DM} m_{\rm DM}}{4\pi \delta}.}\label{lam-eta}
\eea

Noting that the effective fine structure constant $A_{\rm DM}$ depends on the sum of dark matter spins, we need to average the Hulthen self-scattering cross sections over spin states.
For $s$-wave dominance in the dark matter self-scattering, the space-antisymmetric contributions to the self-scattering cross section vanish, so the momentum transfer cross sections are given by
\bea
\sigma^{\rm Hulthen}_{S,T} &\simeq& \frac{8\pi \sin^2\delta^S}{k^2}, \label{HS} \\
\sigma^{\rm Hulthen}_{\chi, T} &\simeq &\frac{2\pi\sin^2\delta^\chi_{s=0}}{k^2}, \label{Hchi} \\
\sigma^{\rm Hulthen}_{X,T} &\simeq & \frac{8\pi}{9k^2} \Big(5\sin^2\delta^X_{s=2}+\sin^2\delta^X_{s=0} \Big) \label{HX}
\eea
where $\delta^S, \delta^\chi_{s=0}, \delta^X_{s=0}, \delta^X_{s=2}$ are the phase shifts given in eq.~(\ref{phaseshift}) with a set of parameters, $m_{\rm DM}=m_S, m_\chi, m_X, m_X$ and
\bea
A_{\rm DM}=A_S,\quad A_\chi, \quad A_X, \quad A_X + \frac{c^2_X m^2_X}{\Lambda^2} \Big(\frac{m_G}{m_X} \Big)^2,
\eea
respectively.

We first note that there is a pole of the gamma function at $\lambda_-=-n$ in the phase shift in eq.~(\ref{phaseshift}), with $n$ being non-negative integer  for which $\delta_0\to  \frac{\pi}{2}$ so the self-scattering cross section is enhanced by $\sigma_{\rm self}\propto 1/v^2$ from eq.~(\ref{diffcross}). In this case, for $k\ll 1$, the resonance condition is given by $1-\sqrt{A_{\rm DM} m_{\rm DM}/(4\pi \delta)}=-n$, namely,
\bea
\frac{3}{2\pi^3} \,\frac{A_{\rm DM}m_{\rm DM}}{m_G} =(n+1)^2,\quad n=0,1,2,\cdots. \label{resonance}
\eea 
For instance, in the limit of $m_G\ll m_{\rm DM}$, for dark matter of arbitrary spin, we have $A_{\rm DM}\simeq \frac{2c^2_{\rm DM}m^2_{\rm DM}}{3\Lambda^2}$, for which the above resonance conditions become
\bea
m_G=\frac{c^2_{\rm DM}}{\pi^3 n^2}\, \frac{ m^3_{\rm DM}}{\Lambda^2}.
\eea
The result leads to an intriguing relation between the DM and spin-2 particle masses for forming the DM bound states. 

We remark that the two-body scattering process for dark matter is typically dominant for the structure formation in galaxies and galaxy clusters. But, if the bound-state formation is not possible for the two-body scattering, there might be a possibility of three-body or many-body bound-state formations in the presence of local substructures with large dark matter densities. But, we don't pursue this possibility in our work, because we focus on the two-body scattering process for solving the small-scale problems as galaxies.

On the other hand, away from the poles of the gamma function, we can make a low-momentum expansion of the phase shift for $v\ll 1$ in the effective range theory, as follows,
\bea
k\cot\delta_0=-\frac{1}{a} + \frac{1}{2}\, r_0 k^2.
\eea
Here, we identify the scattering length and the effective range \cite{effrange}, respectively, as
\bea
a&=&\frac{1}{\delta} \bigg( \psi^{(0)}(1+\eta) + \psi^{(0)}(1-\eta)+2\gamma\bigg) ,\\
r_0&=& \frac{2}{3} a  -\frac{1}{3\delta \eta} \Big[ \psi^{(0)}(1+\eta) + \psi^{(0)}(1-\eta)+2\gamma \Big]^{-2}\nonumber \\
&&\quad \times \bigg[ 3\Big( \psi^{(1)}(1+\eta) - \psi^{(1)}(1-\eta)\Big)+ \eta  \Big( \psi^{(2)}(1+\eta) + \psi^{(2)}(1-\eta)+16\zeta(3) \Big)\bigg]
\eea
where $\gamma\simeq 0.5772$ is the Euler-Mascheroni constant, $\psi^{(n)}(z)$ are the polygamma functions of order $n$, and $\zeta(3)$ is the Riemann zeta function.
Then, the $s$-wave self-scattering cross section is given by
\bea
\sigma^{\rm Hulthen}_{T}=  \frac{4\pi a^2}{1+k^2 (a^2-a r_0)+\frac{1}{4}a^2 r^2_0 k^4}. \label{effrange}
\eea
In this case, due to the presence of the effective range parameter $r_0$ in the velocity expansion, there appears a bound state with binding energy, $E_b=-\frac{k^2}{m_{\rm DM}}$, at the pole with imaginary $k=i r^{-1}_0 (1-\sqrt{1-2r_0/a})$.

For the Born approximation, we can ignore the effective range term for $k\ll 1$ to get the self-scattering cross section as
\bea
\sigma^{\rm Hulthen}_T\simeq \frac{4\pi}{\delta^2}\,\Big[2\gamma+\psi^{(0)}(1+\eta)+\psi^{(0)}(1-\eta)\Big]^2.
\eea
Then, in the non-relativistic limit for dark matter velocity and in the perturbative limit with $A_{\rm DM}m_{\rm DM}/(4\pi m_G)\ll 1$, we can match the approximate results to the Born approximations given in eqs.~(\ref{born-S}), (\ref{born-F}) and (\ref{born-V}).
Consequently, we can extend the region of validity beyond the Born approximations with the Hulth\'en potential, with the following replacements for the self-scattering cross sections,
\bea
\sigma_{S,T}&\simeq&\frac{\sigma^{\rm Hulthen}_{S,T}}{2(\psi^{(2)}(1))^2(6/\pi^2)^4},  \label{exact-S} \\
\sigma_{\chi,T} &\simeq &\frac{3}{8}\cdot  \frac{2\sigma^{\rm Hulthen}_{\chi,T}}{(\psi^{(2)}(1))^2(6/\pi^2)^4}, \label{exact-F}  \\
\sigma_{X,T} &\simeq &\frac{(32-56r_X+27 r^2_X)}{3(4-r_X)^2} \cdot \frac{3\sigma^{\rm Hulthen}_{X,T}}{4(\psi^{(2)}(1))^2(6/\pi^2)^4}. 
\label{exact-V}
\eea
Therefore, we will use the above results to make the numerical analysis for the parameter space for the large self-scattering cross sections in the next section.

\subsection{Sommerfeld effects for dark matter annihilation}

We also discuss the effects of the massive spin-2 mediator on the dark matter annihilations through the Sommerfeld enhancement, as for the non-perturbative self-scattering process in the previous section.

As in the previous section, we replace the Yukawa potential by the Hulth\'en potential and consider the approximate analytic solutions to obtain the Sommerfeld factors.
Following the details in appendix C, we can read the Sommerfeld enhancement factor from the wave function of dark matter at the origin \cite{cassel} as
\bea
S_l &=& \left|\frac{\Gamma(a)\Gamma(b)}{\Gamma(l+1+2iw)}\frac{1}{l!} \right|^2  \label{Sommerfeld}
\eea
where 
\bea
a&=& l+1+iw\Big( 1-\sqrt{1-x/w}\Big), \\
b&=& l+1+iw\Big( 1+\sqrt{1-x/w}\Big),
\eea
with $w=\frac{k}{\delta}$ and $x=\frac{A_{\rm DM}}{4\pi v}$.

For $l=0$, using $|\Gamma(1+i\alpha)|^2=\pi \alpha/\sinh(\pi\alpha)$, we obtain the Sommerfeld factor as follows,
\bea
S_0 = \frac{\frac{\pi}{2}x\,\sinh(2\pi w)}{ \sinh\Big[\pi w\Big(1-\sqrt{1-\frac{x}{w}}\Big)\Big] \sinh\Big[\pi w\Big(1+\sqrt{1-\frac{x}{w}}\Big)\Big]}.
\eea
Then, from  $w=\frac{k}{\delta}=\frac{6}{\pi^2} \frac{m_{\rm DM} v}{m_G}$, the above becomes
\bea
S_0 = \frac{\frac{A_{\rm DM}}{8v}\sinh\Big(\frac{12 m_{\rm DM}v}{\pi m_G}\Big)}{ \sinh\Big[\frac{6 m_{\rm DM}v}{\pi m_G}\Big(1-\sqrt{1-\frac{\pi A_{\rm DM} m_G}{24\, m_{\rm DM}v^2}}\Big)\Big] \sinh\Big[\frac{6 m_{\rm DM}v}{\pi m_G}\Big(1+\sqrt{1-\frac{\pi A_{\rm DM}m_G}{24\, m_{\rm DM}v^2}}\Big)\Big]}. \label{Sommerfeld-s}
\eea
As a result, the $s$-wave  annihilation cross section for dark matter   can be replaced by
\bea
(\sigma_{\rm ann} v)= S_0\,(\sigma^0_{\rm ann}v) \label{cross-SB}
\eea
where $(\sigma^0_{\rm ann}v)$ is the $s$-wave annihilation cross section without Sommerfeld enhancement, namely, the one obtained from the Born limit.
The enhanced annihilation cross sections for higher partial waves  can be similarly obtained from eq.~(\ref{Sommerfeld}).

As discussed for the dark matter self-scattering in the previous subsection, the effective fine structure constant $A_{\rm DM}$ depends on the sum of dark matter spins through $(m_G/m_{\rm DM})^2$ corrections, so does the Sommerfeld factor that we discussed above. 
But, as far as $m_G\ll m_{\rm DM}$ for a sizable Sommerfeld effect, we can ignore the spin dependence in the following discussion.

In the presence of the light spin-2 mediator, dark matter annihilates into a pair of massive spin-2 mediators.
For completeness, we list the corresponding annihilation cross sections at tree level \cite{GMDM1,GMDM2,GLDM}, as follows, 
\bea
(\sigma^0_{\rm ann}v)_{SS\rightarrow GG} &=&\frac{4 c_{S}^4 m_{S}^2}{9 \pi \Lambda^4 }
\frac{(1-r_S)^\frac{9}{2}}{r^4_S  (2-r_S)^2},  \\
 (\sigma^0_{\rm ann}v)_{\chi{\bar\chi}\rightarrow GG} &=&\frac{c_{\chi}^4 m_{\chi}^2}{16 \pi \Lambda^4 }
\frac{(1-r_\chi)^\frac{7}{2}}{r^2_\chi (2-r_\chi)^2}, \\
 (\sigma^0_{\rm ann}v)_{XX\rightarrow GG} &=&\frac{c_{X}^4 m_{X}^2}{324 \pi \Lambda^4 }
\frac{\sqrt{1-r_X}}{r^4_X  (2-r_X)^2} \,
\bigg(176+192 r_X+1404 r^2_X-3108 r^3_X \nonumber \\
&&+1105 r^4_X+362 r^5_X+34 r^6_X \bigg). 
\eea
Thus, all the above annihilation cross sections are $s$-wave, so we can apply the Sommerfeld-enhanced cross sections  according to eq.~(\ref{cross-SB}).
On the other hand, the dark matter annihilation cross section into a pair of SM particles depend on the spin of dark matter and the particles that dark matter annihilate into. In particular, for the dark matter annihilations into a pair of SM fermions, the corresponding annihilation cross sections are $d$-wave, $p$-wave and $s$-wave for scalar, fermion and vector dark matter \cite{GMDM1,GMDM2,diphoton,GLDM}, respectively. Since dark matter annihilation cross sections, in particular, for vector dark matter, are enhanced at a low velocity of dark matter, there are strong indirect constraints from Cosmic Microwave Background at the recombination epoch and the gamma-ray searches from Fermi-LAT at present.

In the limit of $w\ll 1$ or  $m_{\rm DM} v\ll \frac{\pi}{12} \, m_G$, the kinetic energy becomes zero and a pair of dark matter can form a bound state. In this case, the Sommerfeld factor in eq.~(\ref{Sommerfeld}) becomes
\bea
S_l\simeq  \left|\frac{\Gamma(l+1+\sqrt{\omega x})\Gamma(l+1-\sqrt{\omega x})}{l!\Gamma(l+1)}\right|^2.
\eea
Then, there appear resonances at $\omega x=(l+1+n)^2$ for a non-negative integer $n$, leading to the resonance conditions for the $s$-wave Sommerfeld factor, $S_0$, which are the same as in the resonance condition for dark matter self-scattering from eq.~(\ref{resonance}).

\subsection{Comments on the Coulomb limit}

Several remarks on the Coulomb limit of the Sommerfeld factor in eq.~(\ref{Sommerfeld}) are in order.
In the Coulomb limit, taking $m_G\rightarrow 0$ with $w\gg 1$ or  $m_{\rm DM} v\gg \frac{\pi}{12} \, m_G$, the $s$-wave Sommerfeld factor in eq.~(\ref{Sommerfeld-s}) becomes Coulomb-like as   
\bea
S_0\approx  \frac{\pi x}{1-e^{-\pi x}}=\frac{A_{\rm DM}}{4v} \,\frac{1}{1-e^{- A_{\rm DM}/(4v)}}. 
\eea
Here, we note that the approximation with  $m_{\rm DM} v\gg \frac{\pi}{12} \, m_G$ is that the de Broglie wavelength of dark matter is shorter than the radius of graviton exchange, for which the mediator induces a long-range interaction between a pair of dark matter.
For high partial waves with $l>0$, in the limit of $w\gg 1$ with eq.~(\ref{Sommerfeld}), the Coulomb limit of the corresponding Sommerfeld factor \cite{cassel} is given by
\bea
S_l\approx S_0\times \prod^l_{b=1} \bigg(1+\frac{x^2}{4b^2} \bigg)=S_0\times \prod^l_{b=1}  \bigg(1+\frac{A^2_{\rm DM}}{64\pi^2  b^2v^2} \bigg).
\eea
As the perturbative cross sections of higher partial waves are suppressed by $v^{2l}$, in the limit of $v\rightarrow 0$, all the higher partial waves have the same velocity dependence through the Sommerfeld factor $S_0$.
But, for $v\sim \frac{\pi}{12}\, m_G/m_{\rm DM}$, the Sommerfeld enhancement factor saturates to a constant value for $s$-wave or reaches a maximum for $p$-wave.

We need to stress that there is a caution to take the Coulomb limit for the massive spin-2 mediator, due to the discontinuity in the number of degrees of freedom coming from extra polarization states. So, we are not taking the massless limit for the massive spin-2 mediator in our later analysis, but we comment that the non-linear interactions for a light spin-2 mediator can be important due to the extra longitudinal degrees of freedom.  The concrete discussion on non-linear interactions of the massive spin-2 mediator are beyond the scope of our work, but we make a qualitative comment on them in the following. 

The helicity-0 mode $\pi$ in the decoupling limit of massive gravity is described by the cubic Galileon theory \cite{dRGT-review}, with the couplings to dark matter and the SM, in the following,
\bea
{\cal L}_G =\frac{1}{2} (\partial\pi)^2 -\frac{1}{\Lambda^3_3}\, (\partial\pi)^2 \Box \pi -\frac{1}{\Lambda}\,\pi \Big(c_{\rm DM}T_{\rm DM} + c_{\rm SM} T_{\rm SM} \Big)
\eea
where $T_{\rm DM},  T_{\rm SM} $ are the traces of energy-momentum tensors for dark matter and the SM, respectively, and the strong coupling scale is given by $\Lambda_3=(m^2_G\Lambda/c_{\rm DM})^{1/3}$ for $c_{\rm DM}>c_{\rm SM}$.  Then, the helicity-0 mode contribution gets suppressed below the Vainshtein radius, $r_{*}$, given by
\bea
r_{*}= \frac{1}{\Lambda_3} \Big(\frac{c_{\rm DM} m_{\rm DM}}{4\pi\Lambda}\Big)^{1/3}=m^{-1}_G \Big(\frac{m_G}{m_{\rm DM}} \Big)^{1/3} \Big(\frac{3A_{\rm DM}}{8\pi}\Big)^{1/3}. \label{vradius}
\eea
Here, we took the effective fine structure constant by $A_{\rm DM}=\frac{2c^2_{\rm DM} m^2_{\rm DM}}{3\Lambda^2}$.
Therefore,  the non-linear interactions become important at the distances, $r\lesssim r_*$, such that the helicity-0 mode contribution gets suppressed. In this case, the effective potential for dark matter gets smaller than in the Coulomb limit without non-linear interactions included.  This is the case when the de Broglie length of dark matter is smaller than the Vainshtein radius, that is, $1/m_{\rm DM} v\lesssim r_* $,  with eq.~(\ref{vradius}), which amounts to
\bea
\frac{m_{\rm DM} v}{m_G} \gtrsim  \Big(\frac{m_{\rm DM}}{m_G} \Big)^{1/3} \Big(\frac{8\pi}{3A_{\rm DM}}\Big)^{1/3}. \label{vaineff}
\eea
For instance, for $m_G/m_{\rm DM}=\frac{3A_{\rm DM}}{2\pi^3n^2}$ at resonances from eq.~(\ref{resonance}), eq.~(\ref{vaineff}) becomes $\frac{m_{\rm DM} v}{m_G} \gtrsim 5.6 (n/A_{\rm DM})^{2/3}$. 
But, in our work,  we have taken $m_{\rm DM}v\lesssim m_G$ for which the non-perturbative self-scattering is important, the Vainshstein effect can be ignored.  But, for a larger dark matter momentum such as in galaxy clusters, we would need to take into account the Vainshtein effects due to the non-linear interactions of the massive spin-2 mediator.
We also remark that there is a similar Vainshtein effect on the interactions of the massive spin-2 mediator to the SM, but at much shorter distances than the one for dark matter for $c_{\rm SM}\ll c_{\rm DM}$.

\subsection{Classical regime}

For completeness, in the non-perturbative regime with $A_{\rm DM} m_{\rm DM}/(4\pi m_G)\gtrsim 1$, we also summarize the classical regime with $m_{\rm DM} v\gtrsim  m_G$.
In this case, the Born approximation with $s$-wave only breaks down, so we need to include higher partial waves in the scattering amplitude.
As a result, assuming the effective Yukawa-type potential, $V=-\frac{A_{\rm DM}}{4\pi r} \, e^{-m_Gr}$, to be dominant, the momentum transfer cross section has the following velocity-dependences \cite{classregime},
\bea
\sigma^{\rm cl}_T = \left\{ \begin{array}{ccc} \frac{2\pi}{m^2_G}\,\beta^2 \ln(1+\beta^{-2}),\quad \beta\lesssim 10^{-2},  \vspace{0.2cm} \\ \frac{7\pi}{m^2_G}\,\frac{\beta^{1.8} +280 (\beta/10)^{10.3}}{1+1.4\beta+0.006 \beta^4+160(\beta/10)^{10}}, \quad 10^{-2}\lesssim \beta \lesssim 10^2,  \vspace{0.2cm} \\ \frac{0.81\pi}{m^2_G}\,
\Big(\ln\beta+1 -\frac{1}{2}\ln^{-1}\beta\Big), \quad\beta\gtrsim 10^2  \end{array} \right. \label{classical}
\eea
where $\beta$ is the ratio of the potential energy at $r\sim m^{-1}_G$ to the kinetic energy of dark matter, given by
\bea
\beta\equiv \frac{A_{\rm DM} m_G}{2\pi m_{\rm DM} v^2}.
\eea
This regime is called the strongly-coupled regime, because of a large $A_{\rm DM}$ for a sizable $\beta$.

In general, the results with higher partial waves in the classical regime are subject to the momentum-dependent interactions of the spin-2 mediator. Thus, comments on the validity of the momentum expansion in our case are in order. As illustration, we consider the momentum expansion of the tree-level scattering amplitude for scalar dark matter, $SS\to SS$, as follows,
\bea
{\cal M}(SS\to SS)&\simeq&\frac{4 A_S m^2_S }{{\vec q}^2+m^2_G} \bigg[ 1+\frac{3}{2} (v^\perp)^2+\frac{3}{8} (v^\perp)^4 \nonumber \\
&&\quad +\frac{{\vec q}^2}{4m^2_S}\Big( 1+\frac{3}{2} (v^\perp)^2\Big)-\frac{{\vec q}^4}{4m^2_S m^2_G} \Big(1-\frac{m^2_G}{4m^2_S} \Big) \bigg]
\eea
where $A_S=\frac{2c^2_S m^2_S}{3\Lambda^2}$ and ${\vec q}$ is the momentum transfer between dark matter particles, and ${\vec v}^\perp$ is the velocity component satisfying ${\vec v}^\perp\cdot {\vec q}=0$ with $(v^\perp)^2={\vec v}^2-\frac{{\vec q}^2}{m^2_S}$ for ${\vec v}$ being the relative velocity between dark matter particles.
For $(v^\perp)^2, \,\, {\vec q}^2/m^2_S,\, {\vec q}^4/(m^2_Sm^2_G) \ll 1$, the momentum expansion is justified and the effective Yukawa potential in the previous section in eq.~(\ref{sc1}) is recovered. However, for a large momentum transfer, the momentum expansion would break down, so we would need to include higher order terms beyond the tree level. 
Similar conclusions can be obtained for fermion and vector dark matter cases.

Consequently, for $|{\vec q}|\sim m_{\rm DM} |{\vec v}|$, in the semi-relativistic regime with $|{\vec v}|\lesssim 1$, the validity of the momentum expansion with ${\vec q}^4/(m^2_{\rm DM}m^2_G) \lesssim 1$ for the dark matter self-scattering requires
\bea
\frac{m_{\rm DM} v}{m_G}\lesssim \Big(\frac{m_{\rm DM}}{m_G} \Big)^{1/2}, \label{pertbound}
\eea
resulting in
\bea
\frac{m_{\rm DM} v}{m_G}\lesssim \frac{4.5 n}{A^{1/2}_{\rm DM}}, \qquad \beta=\frac{A_{\rm DM} m_{\rm DM}}{2\pi m_G} \Big(\frac{m_G}{m_{\rm DM}v} \Big)^2 \gtrsim  \frac{A_{\rm DM}}{2\pi}
\eea
for $m_{G}/m_{\rm DM}=\frac{3A_{\rm DM}}{2\pi^3n^2}$ at resonances from eq.~(\ref{resonance}).
Therefore, in this case, our approximation with the effective Yukawa potential in eq.~(\ref{classical}) is valid as far as the above bounds are satisfied. Away from the resonances, as far as the mass of the spin-2 mediator is above the resonances, the above bounds  in eq.~(\ref{pertbound}) are still satisfied.

For the later discussion on the self-scattering, we focus on the non-relativistic regime for the dark matter self-scattering, that is, with $m_{\rm DM}v\lesssim m_G$, so the Born regime and the quantum regime discussed in the previous subsections are relevant. Then, our results for the dark matter self-scattering are insensitive to higher partial waves in the classical regime.

\subsection{Velocity-dependent self-scattering: numerical analysis}

We are now in a position to use the results in the previous subsections and discuss the numerical analysis of the self-scattering cross section of dark matter in the parameter space of our model. 
To this, we use the approximate momentum transfer self-scattering cross sections in eqs.~(\ref{exact-S}), (\ref{exact-F})  and (\ref{exact-V}), and compare them to the Born cross sections given in eqs.~(\ref{born-S})-(\ref{born-V}). 
It has been shown that the approximate solution for the Sommerfeld factor coincides with the full numerical solution, except around the resonances where there is a noticeable shift \cite{cassel}. For a larger Sommerfeld factor close to the resonances, the treatment of the full numerical solution would be necessary \cite{cassel}.

\begin{figure}[h]
\centering 
\includegraphics[width=.30\textwidth]{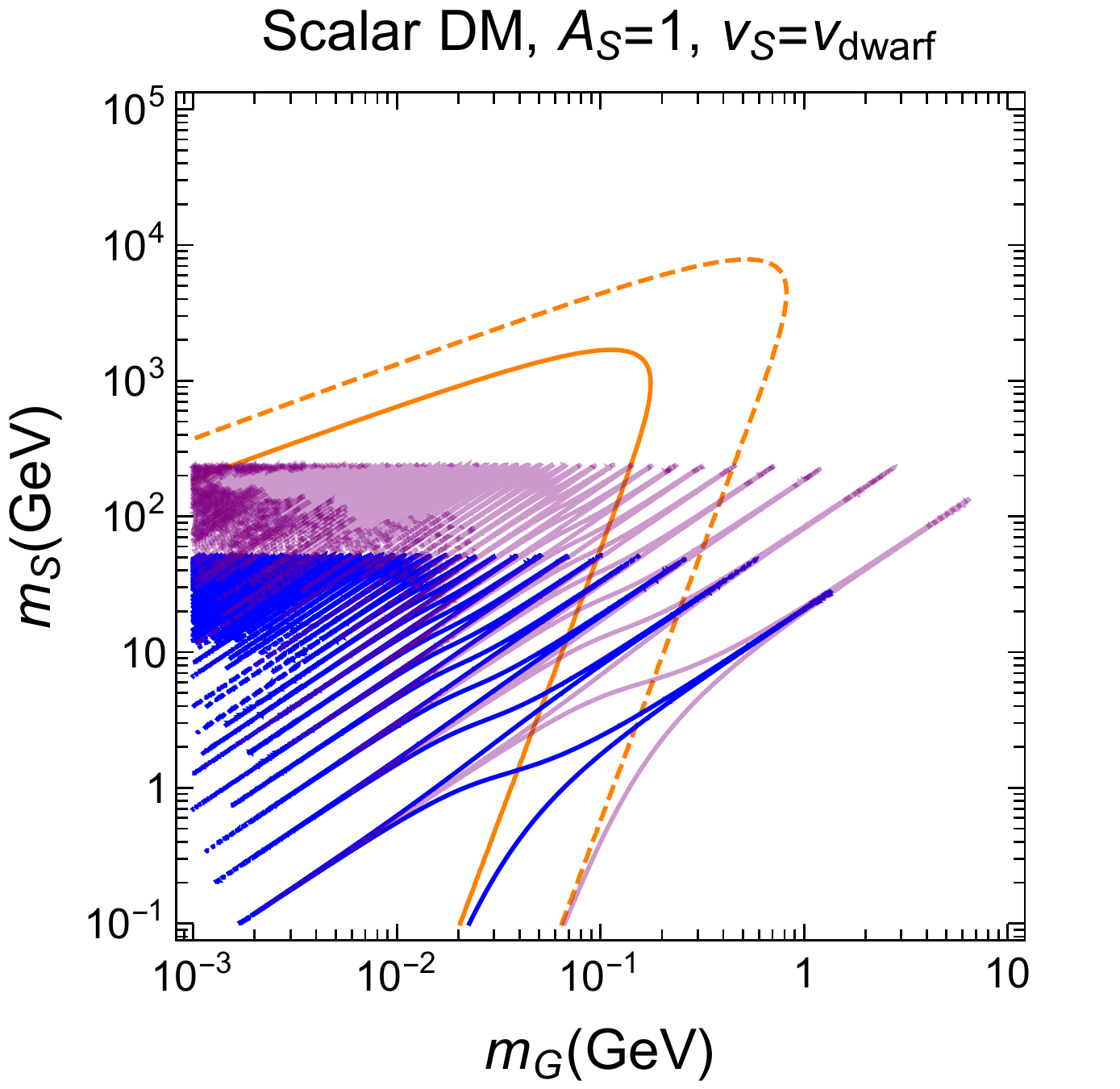} \,\,
\includegraphics[width=.30\textwidth]{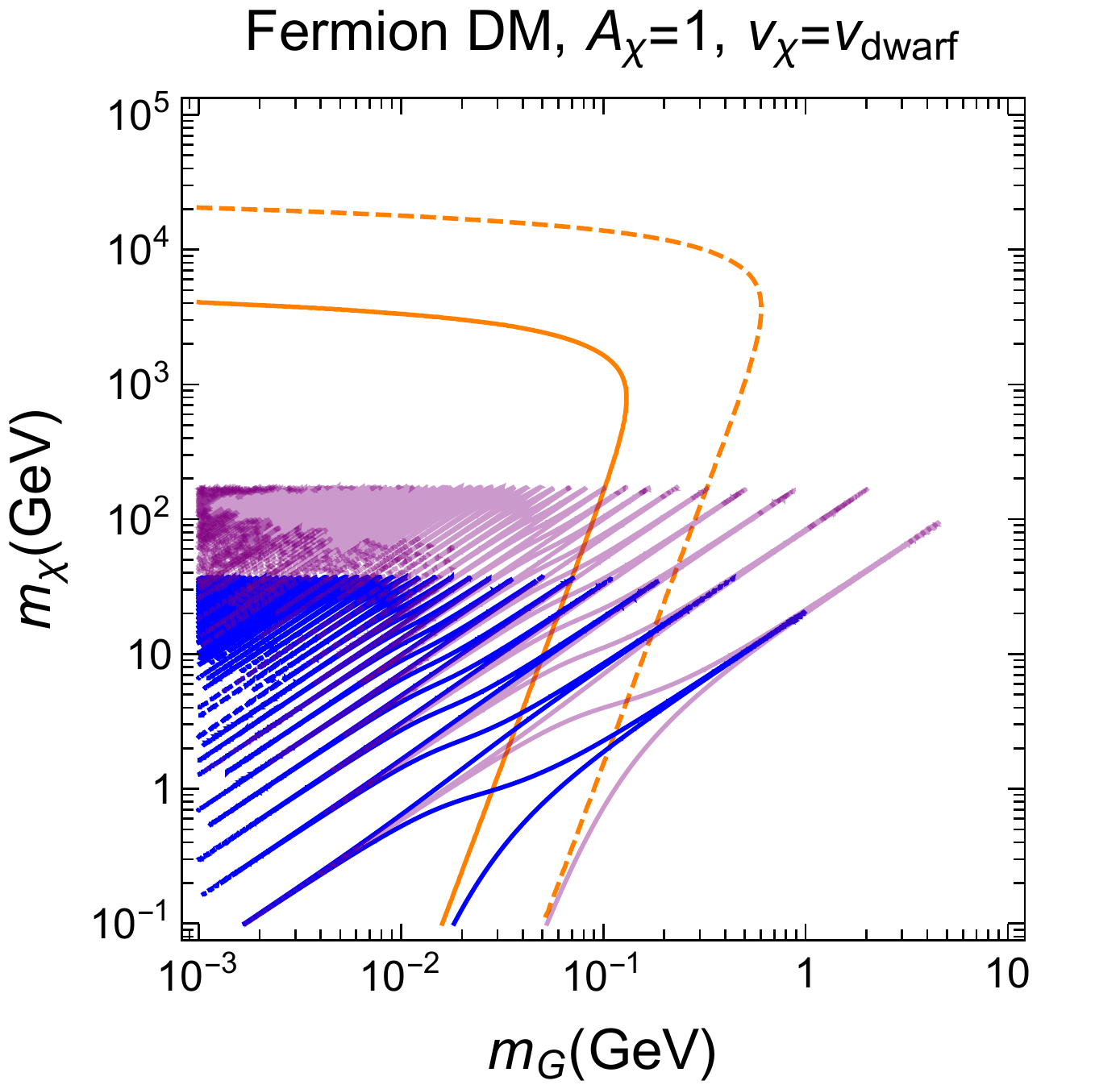} \,\,
\includegraphics[width=.30\textwidth]{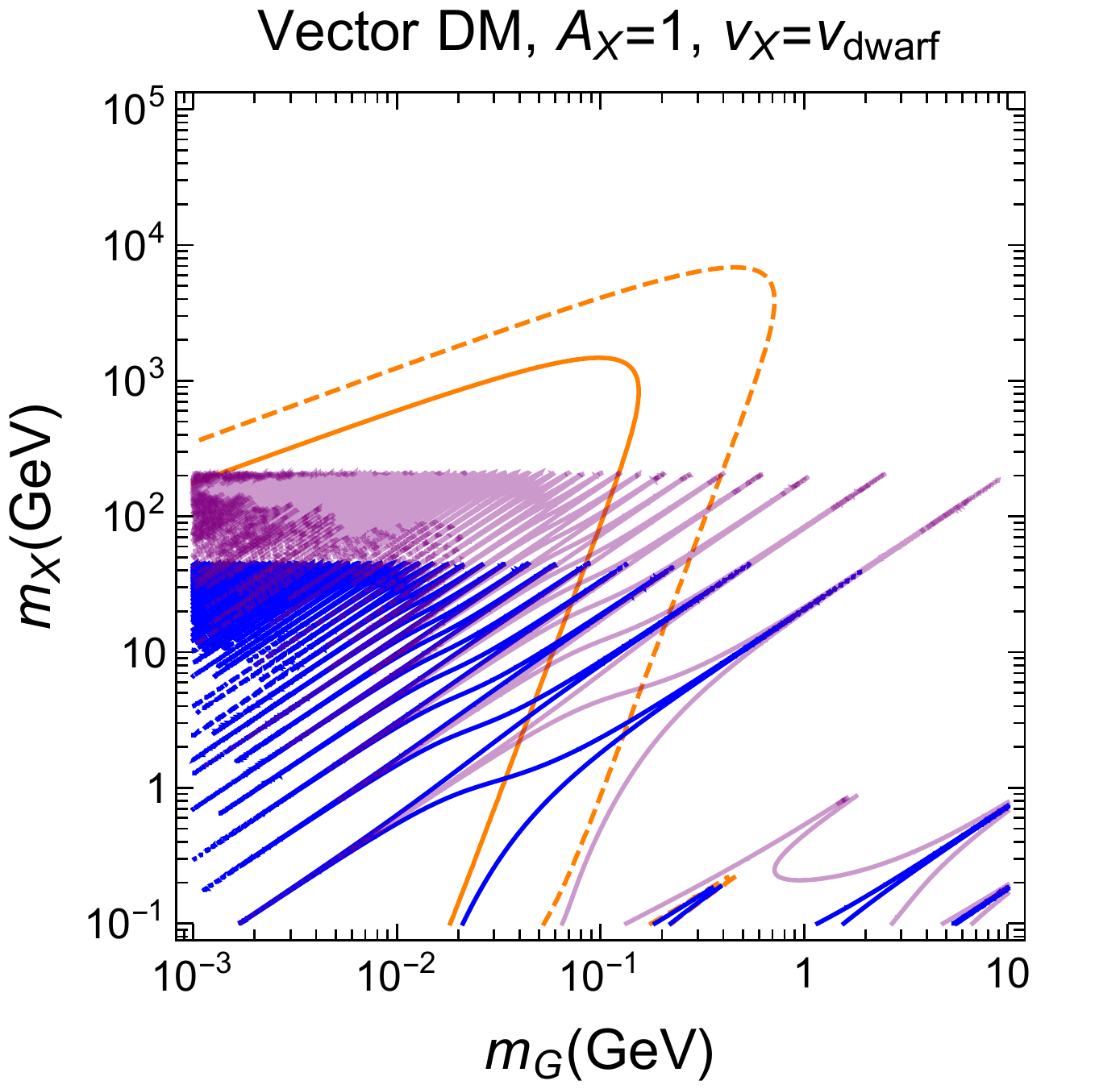}  \vspace{0.3cm} \\
\includegraphics[width=.30\textwidth]{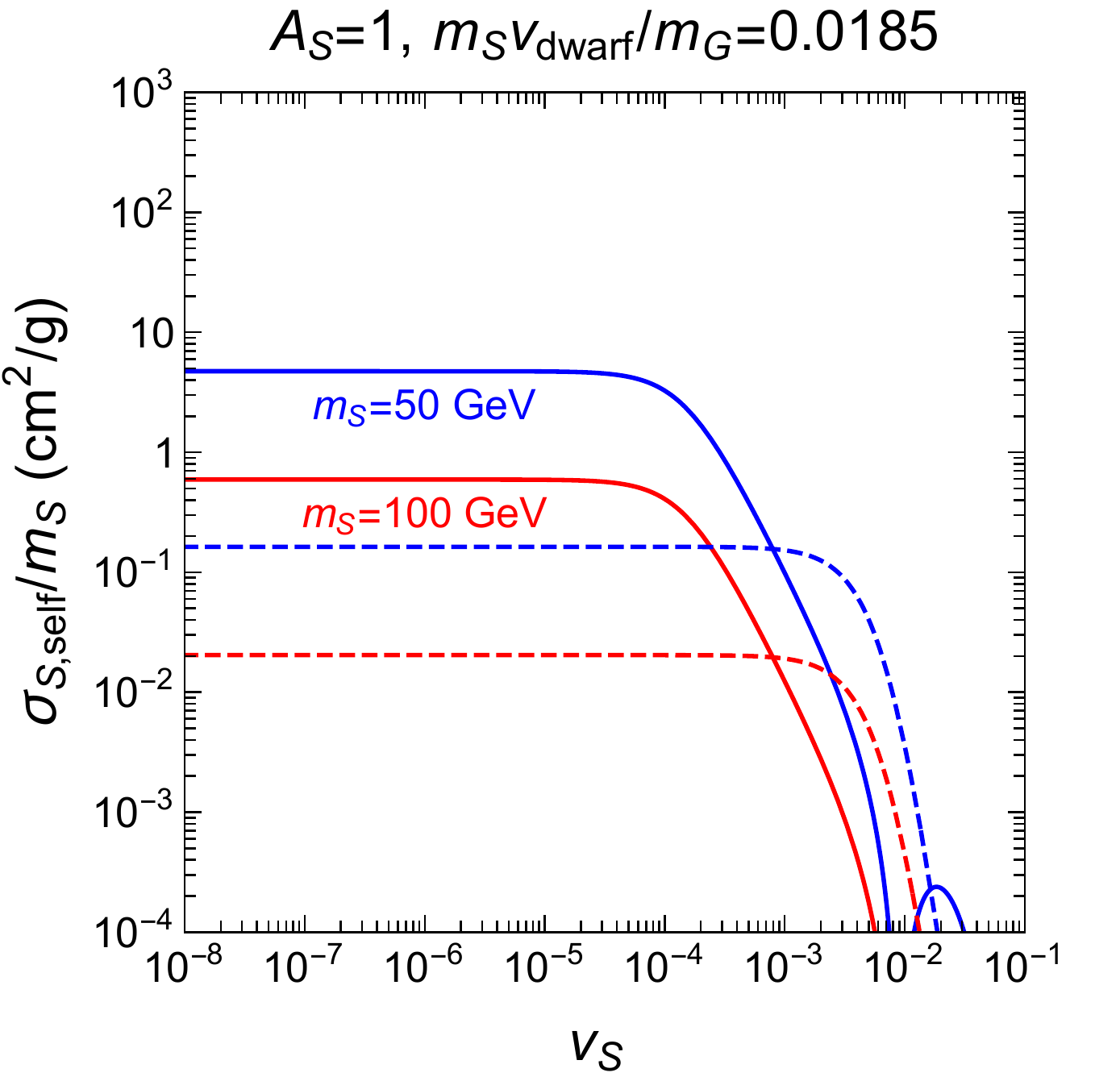} \,\,
\includegraphics[width=.30\textwidth]{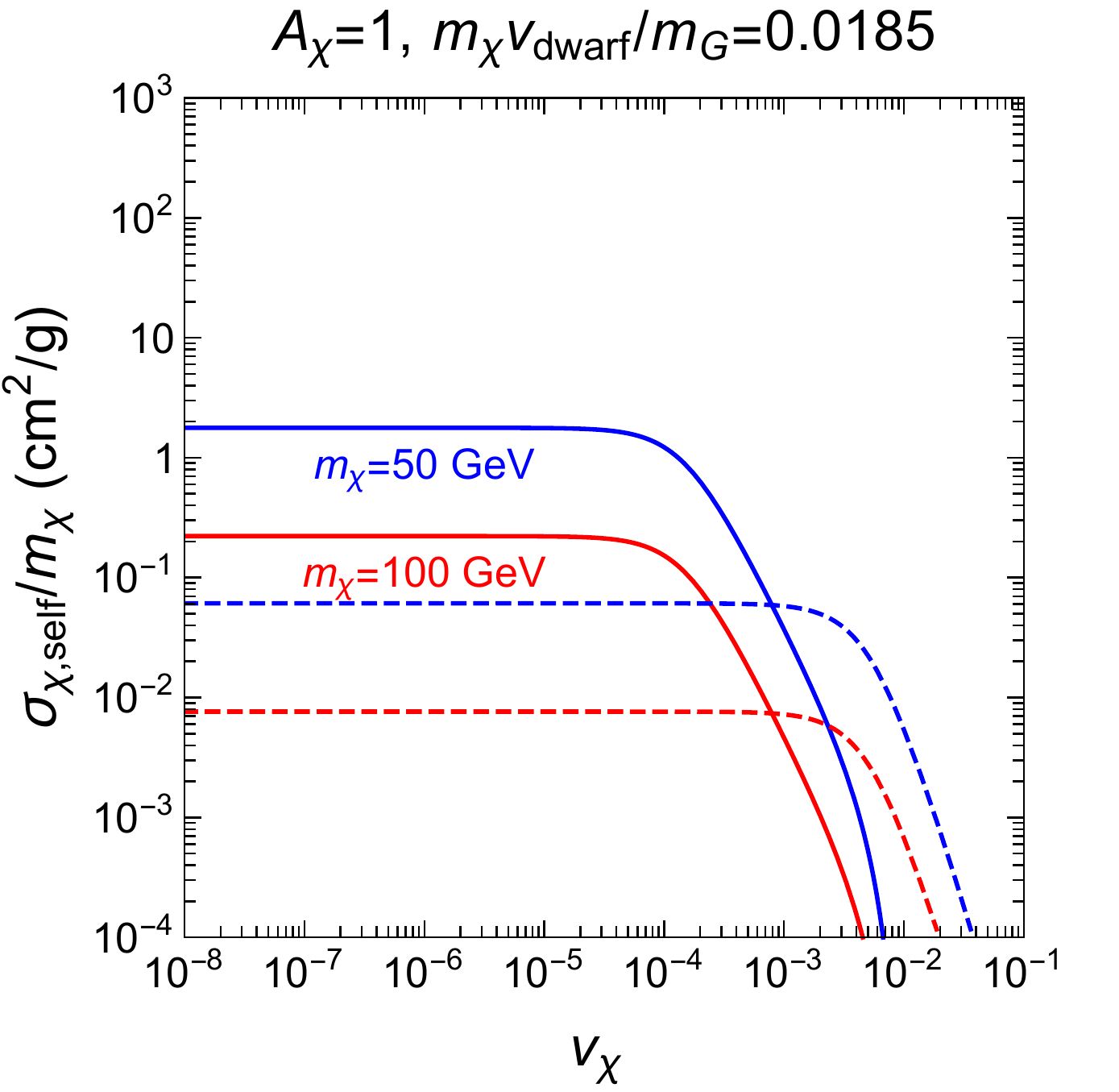} \,\,
\includegraphics[width=.30\textwidth]{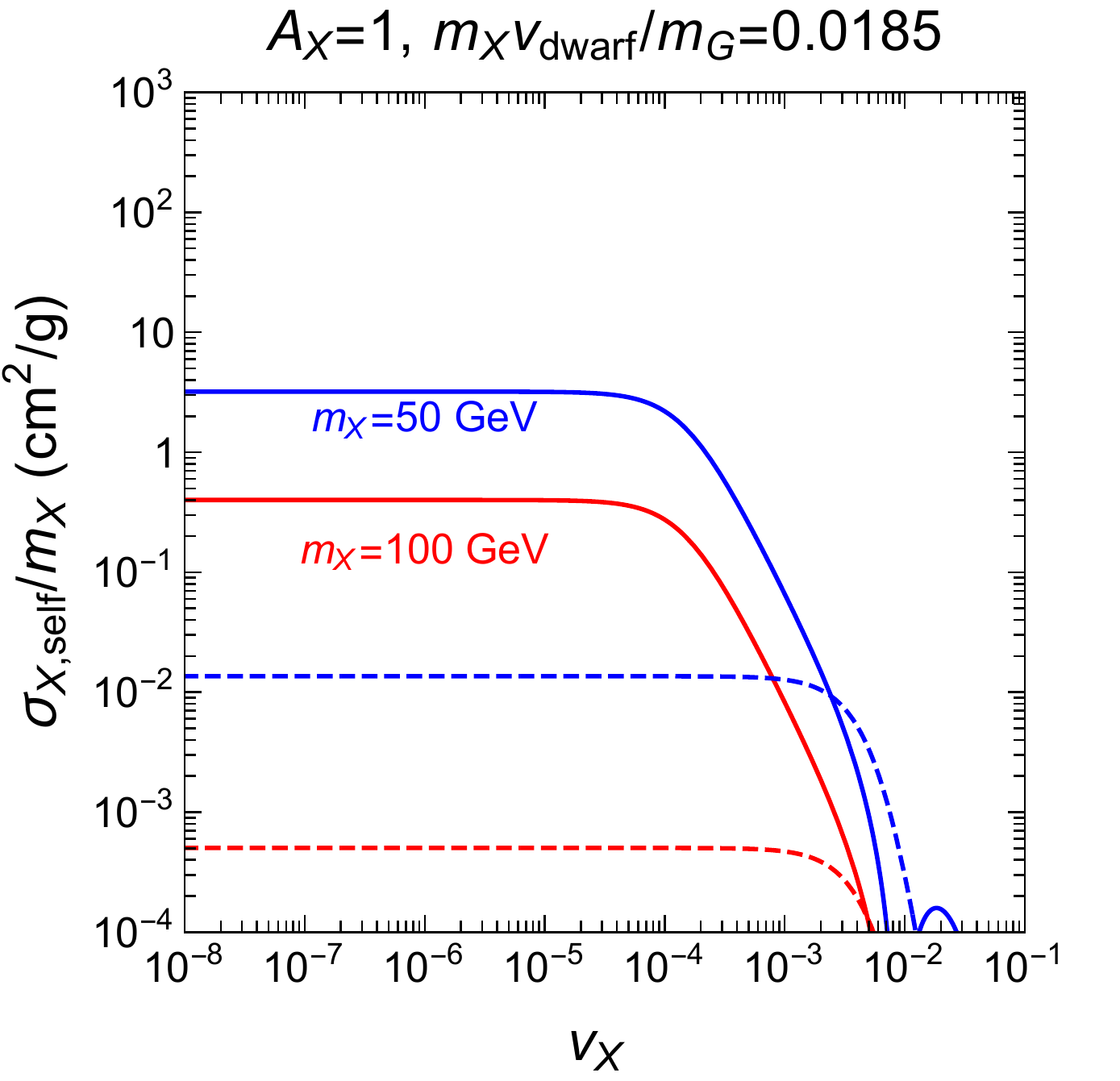} 
\caption{\label{fig:a1}  Top: Contours of the DM self-scattering cross section divided by the DM mass at dwarf galaxies with $v_{\rm dwarf}=10^{-4} c$ in the parameter space for $m_G$ vs $m_{\rm DM}$. We took $\sigma_T/m_{\rm DM}=0.1,10\,{\rm cm^2/g}$. The orange dashed and solid lines are the results for the Born cross section, whereas the purple and blue lines are those for the non-perturbative cross section. Middle (Bottom): The DM self-scattering cross section divided by the DM mass as a function of the DM velocity for $m_{\rm DM}v_{\rm dwarf}/m_G=0.0185$ and two fixed DM masses in each plot. Dashed and solid lines are for the Born and non-perturbative cross sections, respectively.
$A_{\rm DM}=1$ is chosen for all the plots and DM spins are taken to $s=0,1/2,1$ from left to right in each panel. }
\end{figure}

In the top panel of Fig.~\ref{fig:a1}, we depicted the contours in the parameter space for $m_G$ vs $m_{\rm DM}$ for the DM self-scattering cross section divided by the DM mass. We have fixed the DM velocity to $v_{\rm dwarf}=10^{-4} c$  at dwarf galaxies, the effective fine structure constant to $A_{\rm DM}=1$, and the contours are shown for $\sigma_T/m_{\rm DM}=0.1, 10\,{\rm cm^2/g}$. The orange dashed and solid lines indicate the results with the Born cross section. On the other hand, the results with the non-perturbative cross section are shown in purple and blue lines. The cases for scalar, fermion and vector dark matter are shown from left to right in the panel. 
We found that the DM masses up to $200\,{\rm GeV}$ and the spin-2 mediator masses up to $8\,{\rm GeV}$ are required to get the self-scattering cross section for  solving the small-scale problems.
We note that in the Born approximation in orange lines, the results are almost the same for scalar and vector dark matter, while the result for fermion dark matter differ from the other two cases, due to the fact that there are both particle-particle and particle-antiparticle scattering processes. But, the results with the Born cross section deviate significantly from the non-perturbative cross section in the region with a light spin-2 mediator.

In the second panel of Fig.~\ref{fig:a1}, we also showed the DM self-scattering cross section divided by the DM mass as a function of the DM velocity for $A_{\rm DM}=1$ and several choices of the DM and spin-2 mediator masses.
The cases for scalar, fermion and vector dark matter are shown from left to right in each panel. Dashed and solid lines indicate the Born self-scattering cross section and  and the non-perturbative self-scattering cross section from the Hulth\'en potential, respectively. 
In the second panel, we chose $m_{\rm DM}=50, 100\,{\rm GeV}$ and $m_{\rm DM} v_{\rm dwarf}/m_G=0.0185$ (i.e. $m_G=0.27, 0.54\,{\rm GeV}$) for blue and red lines, respectively. In this case, the resulting self-scattering cross section gets saturated to a constant value below $v_{\rm DM}\sim 10^{-4}$ and it becomes highly suppressed at $v_{\rm DM}\sim 10^{-2}$ below the bounds from Bullet cluster \cite{bullet}.
We note that the non-perturbative cross section shows up smaller than the Born cross section in the regime with a low velocity of dark matter, due to higher order corrections.

\begin{figure}[h]
\centering 
\includegraphics[width=.30\textwidth]{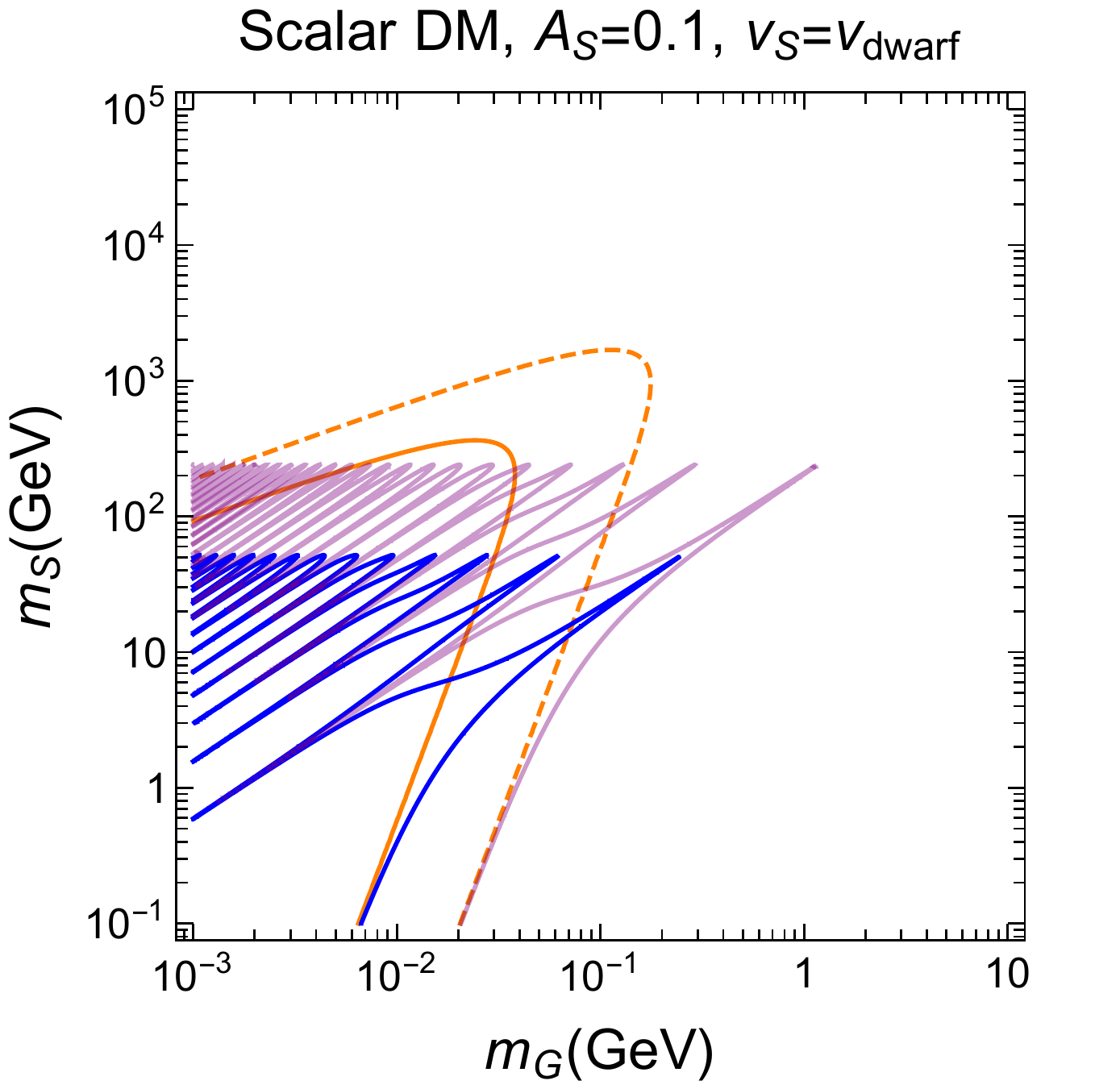} \,\,
\includegraphics[width=.30\textwidth]{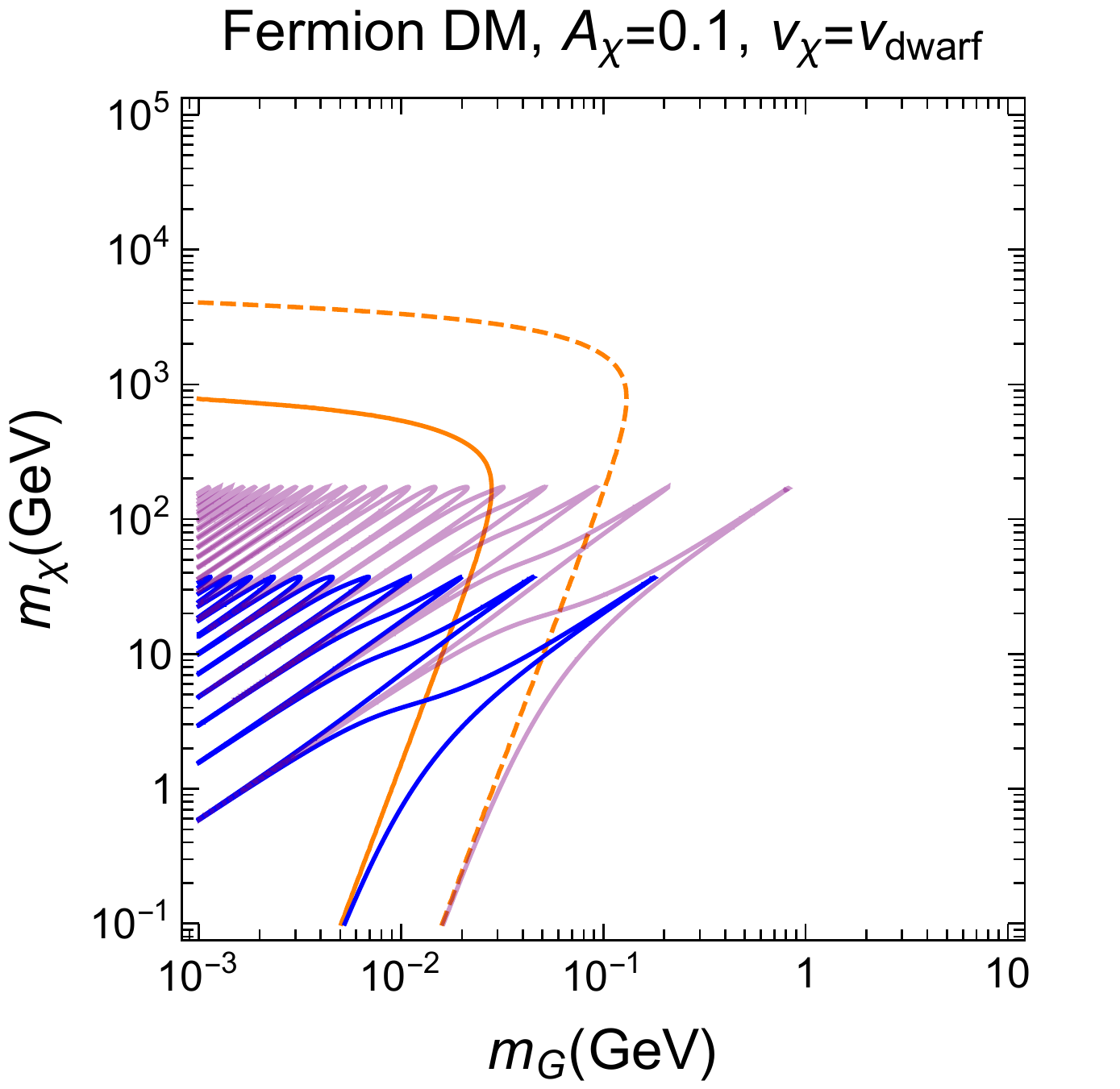} \,\,
\includegraphics[width=.30\textwidth]{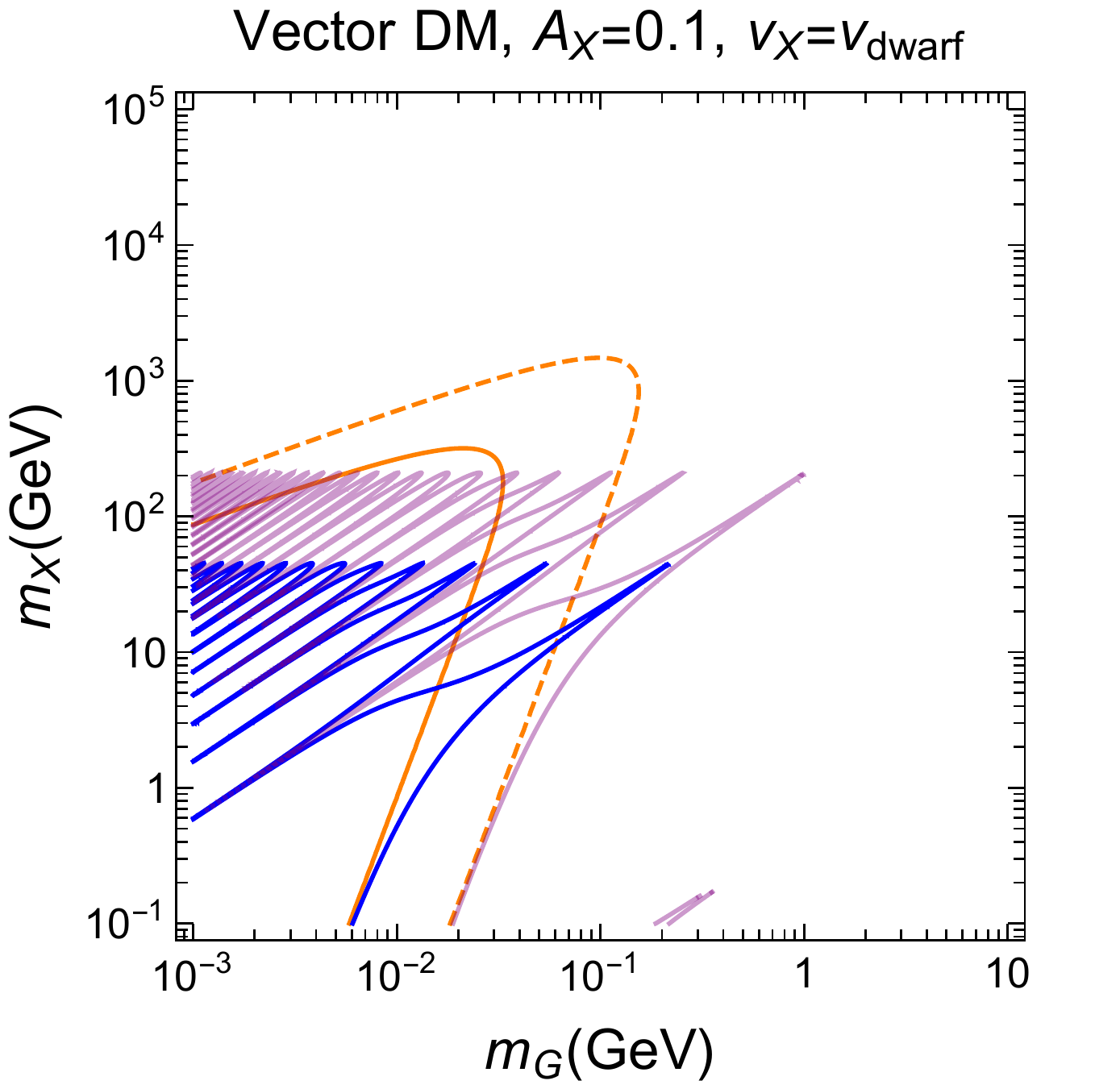}   \vspace{0.3cm} \\
\includegraphics[width=.30\textwidth]{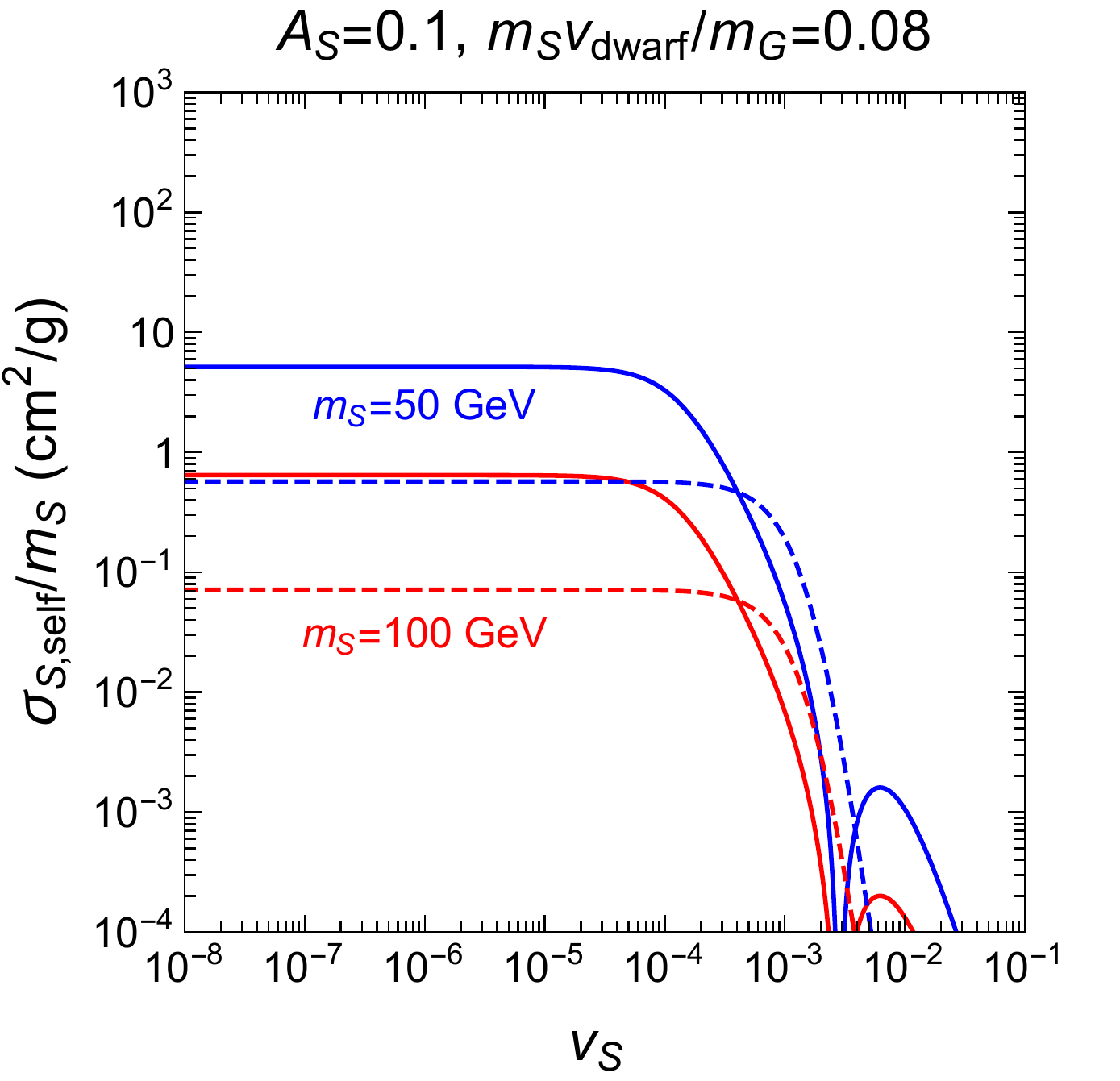} \,\,
\includegraphics[width=.30\textwidth]{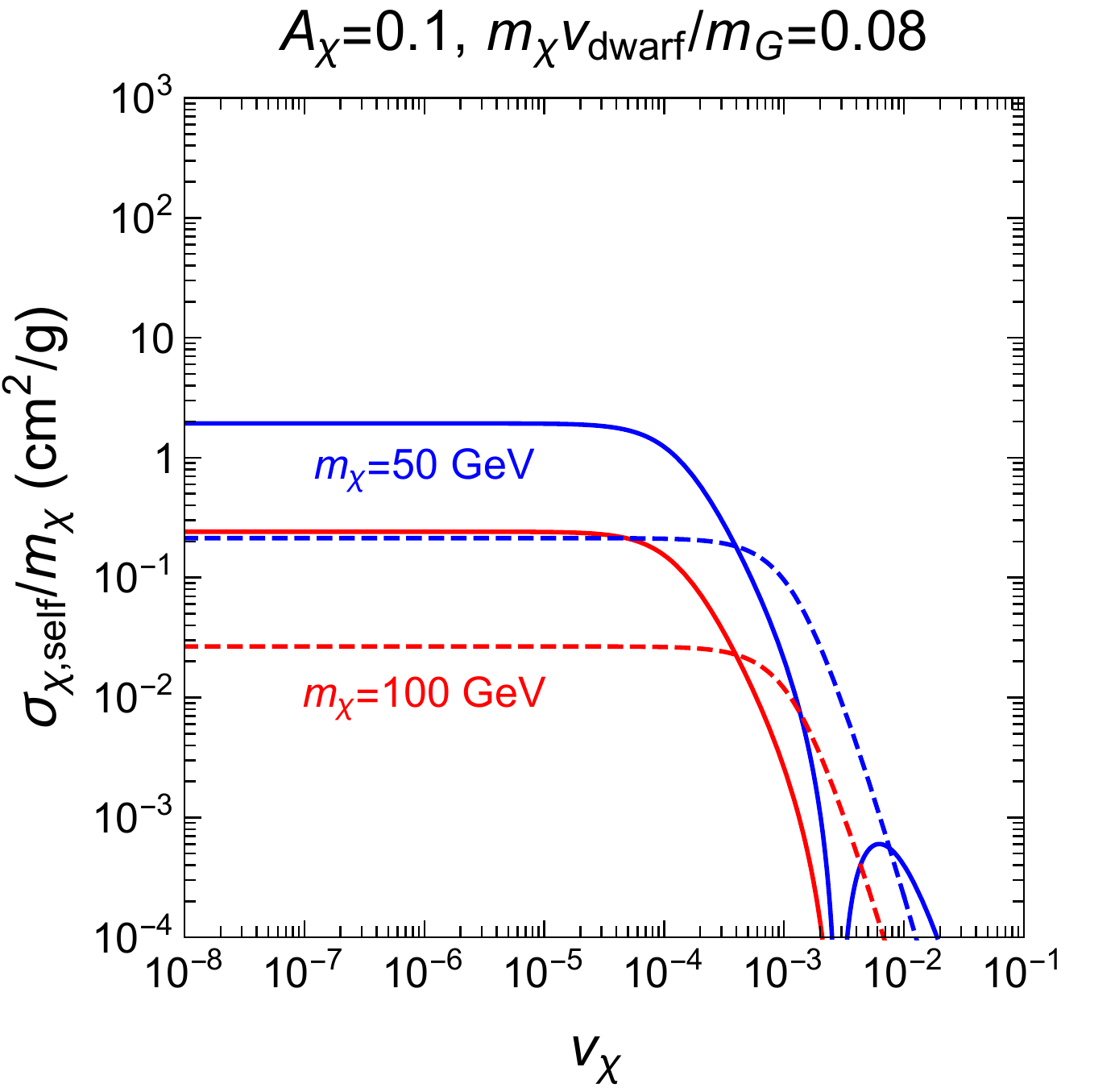} \,\,
\includegraphics[width=.30\textwidth]{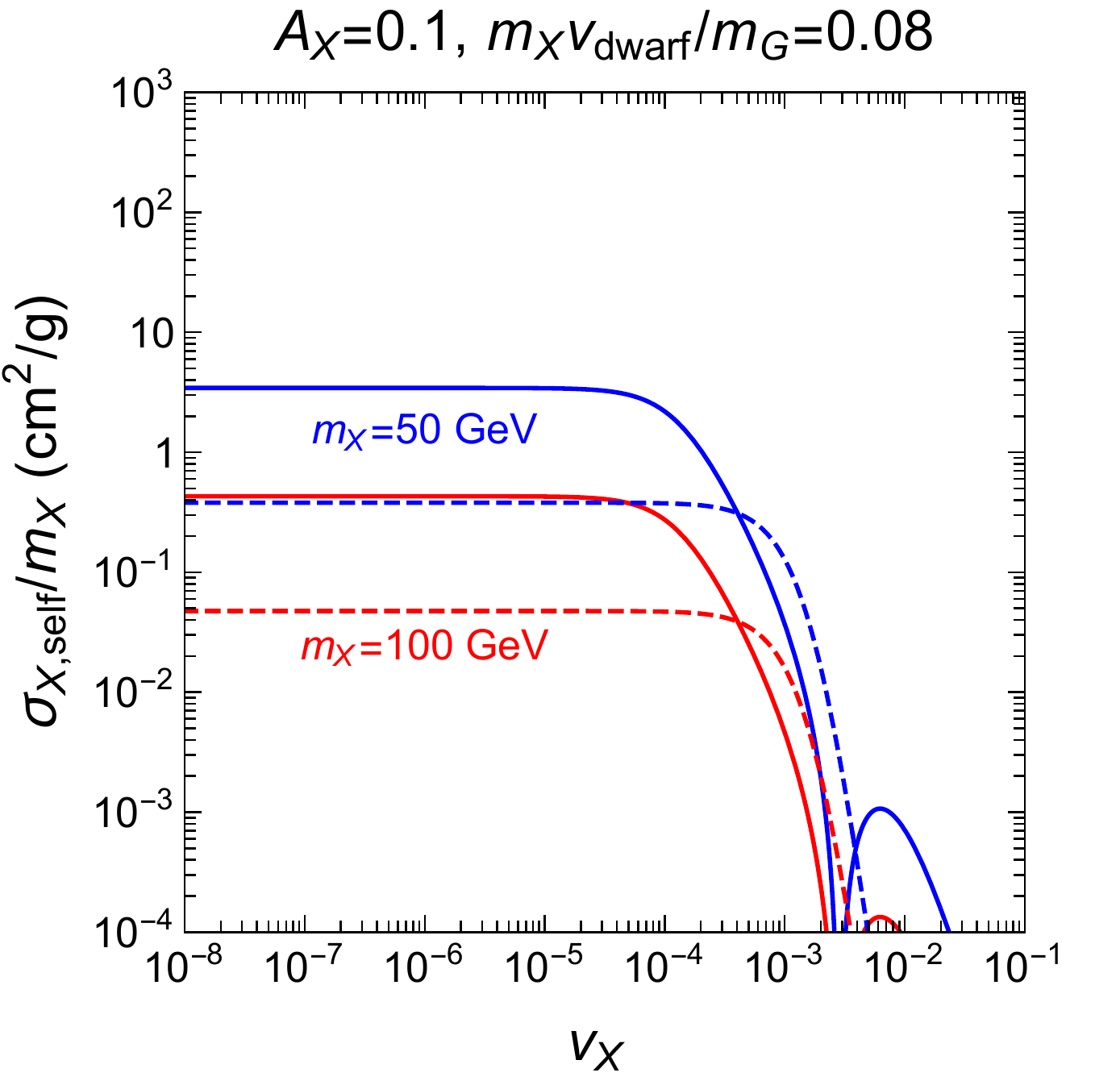} 
\caption{\label{fig:a01}  The same as in Fig.~\ref{fig:a1}, except $A_{\rm DM}=0.1$ for all the plots and $m_{\rm DM}v_{\rm dwarf}/m_G=0.08$ for the plots in the bottom panel. }
\end{figure}

We have repeated the similar analysis of the self-scattering cross section for $A_{\rm DM}=0.1$ in Fig.~\ref{fig:a01}. The cases for scalar, fermion and vector dark matter are shown from left to right in each panel. 
In the top panel of Fig.~\ref{fig:a01}, we showed that the DM masses up to $200\,{\rm GeV}$ and the spin-2 mediator masses up to $1\,{\rm GeV}$ are requires to get the self-scattering cross section for solving the small-scale problems. 
Therefore, we need a lighter spin-2 mediator for the enhanced self-scattering cross section with a smaller value of $A_{\rm DM}$.
On the other hand, in the second panel of Fig.~\ref{fig:a01}, we also drew  the DM self-scattering cross section divided by the DM mass as a function of the DM velocity. We took $m_{\rm DM}=50, 100\,{\rm GeV}$ and $m_{\rm DM} v_{\rm dwarf}/m_G=0.08$ (i.e. $m_G=0.06, 0.12\,{\rm GeV}$) for blue and red lines in the second panel of Figs.~\ref{fig:a01}. 
The saturation behaviors of the self-scattering cross section below galaxy scales are similar to the cases in Fig.~\ref{fig:a1}.

\begin{figure}[h]
\centering 
\includegraphics[width=.40\textwidth]{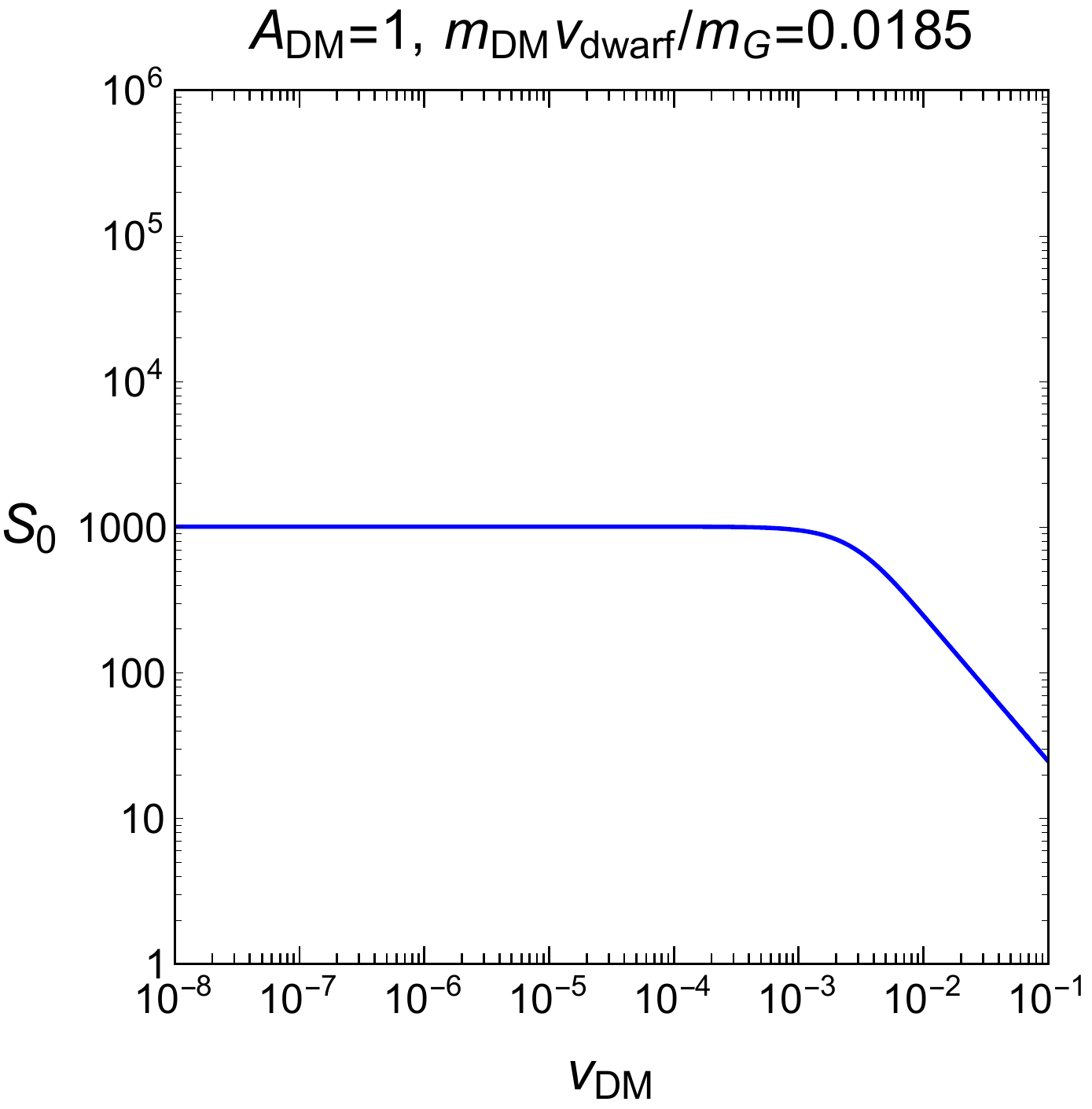} \,\,
\includegraphics[width=.40\textwidth]{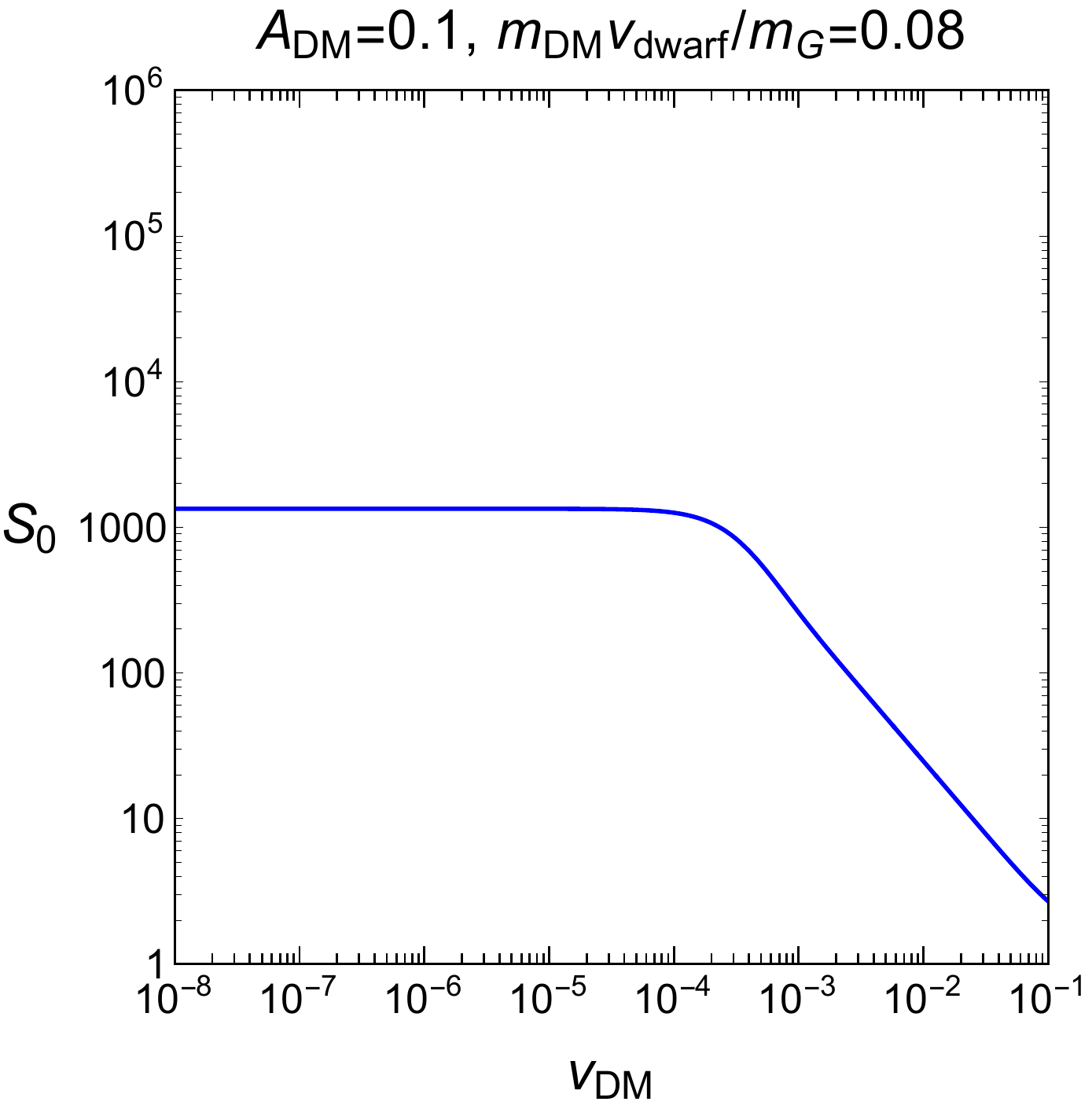}  
\caption{\label{fig:sb}  Sommerfeld factor for dark matter annihilation for $A_{\rm DM}=1, 0.1$ on left and right plots, respectively.  We also took  $m_{\rm DM} v_{\rm dwarf}/m_G=0.0185, 0.08$ on left and right plots, respectively.  }
\end{figure}

Finally, in Fig.~\ref{fig:sb}, we have also shown the Sommerfeld factor in eq.~(\ref{Sommerfeld-s}) for the $s$-wave annihilation of dark matter. We took the parameter space shown in Figs.~\ref{fig:a1} and \ref{fig:a01}: $A_{\rm DM}=1, 0.1$ on left and right plots, respectively. For comparison to the results for self-scattering cross section in Figs.~\ref{fig:a1} and \ref{fig:a01}, e also fixed the ratio of the DM mass to the mass of the spin-2 mediator to $m_{\rm DM} v_{\rm dwarf}/m_G=0.0185, 0.08$ on left and right plots, respectively. Then, the Sommerfeld factor is saturated to a constant value at a low velocity of dark matter, similarly to the case of the non-perturbative cross section. 
The saturated value of the Sommerfeld factor varies between $S_0\sim 10^2$ and $10^3$, depending on $A_{\rm DM}$.  The larger $A_{\rm DM}$, the larger the Sommerfeld factor. The Sommerfeld factor becomes suppressed below $v\sim 10^{-4}$ at dwarf galaxies ($10^{-3}$ at galaxies) for $m_{\rm DM} v_{\rm dwarf}/m_G=0.08 (0.0185)$.  
The above results are closely tied to the similar behavior of the non-perturbative  self-scattering cross section at small velocities.

\section{Conclusions}

We have presented the novel results for the effective potential for the self-scattering of dark matter in the presence of a massive spin-2 mediator. The spin-independent interactions are determined mostly by the mass of dark matter in the limit of a light spin-2 particle,  thus we get  the similar Yukawa potential for all spins of dark matter. 
There are spin-dependent interactions such as spin-spin or dipole-dipole interactions for fermion or vector dark matter, although they are suppressed by the mass of the spin-2 particle or the extra power with $1/r^2$, respectively, as compared to the Yukawa potential. We identified the new forms of spin-dependent effective operators for vector dark matter for the first time. 

Describing the self-scattering process of dark matter in the limit where the effective Yukawa potential is dominant, we not only obtained the self-scattering cross sections for dark matter in the Born limit but also included the non-perturbative effects in order to cover the strong coupling regime. We found that there is a wide range of the parameter space in the DM and spin-2 particle masses where  the small-scale problems at galaxies are resolved and the bounds from galaxy clusters can be satisfied.

\section*{Acknowledgments}

The work is supported in part by Basic Science Research Program through the National Research Foundation of Korea (NRF) funded by the Ministry of Education, Science and Technology (NRF-2019R1A2C2003738 and NRF-2018R1A4A1025334). 
The work of YJK is supported in part by the National Research Foundation of Korea (NRF-2019-Global Ph.D. Fellowship Program).

\def\theequation{A.\arabic{equation}}

\setcounter{equation}{0}

\vskip0.8cm
\noindent
{\Large \bf Appendix A:  Amplitudes for DM self-scattering} 
\vskip0.4cm
\noindent

Here is the summary of derivation of the amplitudes for self-scattering of dark matter.

\noindent
\underline {\bf Fermion dark matter}:

The relevant effective interactions for traceless parts of fermion dark matter are
\bea
16{\tilde T}^\chi_{\mu\nu} {\tilde T}^{\chi,\mu\nu}
&=&(2(p_1+p_2)\cdot (k_1+k_2)) ({\bar u}_\chi(k_2)\gamma_\mu u_\chi(k_1)) ({\bar u}_\chi(p_2)\gamma^\mu u_\chi(p_1)) \nonumber \\
&&- ({\bar u}_\chi(k_2) (\slashed{k}_1+\slashed{k}_2) u_\chi(k_1))({\bar u}_\chi(p_2) (\slashed{p}_1+\slashed{p}_2) u_\chi(p_1))  \nonumber \\
&&+ 2 ({\bar u}_\chi(k_2) (\slashed{p}_1+\slashed{p}_2) u_\chi(k_1))({\bar u}_\chi(p_2) (\slashed{k}_1+\slashed{k}_2) u_\chi(p_1))  \label{tracefree}
\eea
Then, using Dirac equations, $\slashed{p}u_\chi(p)=m_\chi u_\chi(p)  $ and ${\bar u}_\chi(p)\slashed{p}= {\bar u}_\chi(p) m_\chi$ and Gordon identities,
\bea
{\bar u}_\chi (k_2) \gamma^\mu u_\chi(k_1)&=&\frac{1}{2m_\chi} {\bar u}_\chi(k_2) \Big((k_1+k_2)^\mu-i\sigma^{\mu\rho}q_\rho \Big)u_\chi(k_1),   \label{GI1} \\
{\bar u}_\chi (p_2) \gamma^\nu u_\chi(p_1)&=&\frac{1}{2m_\chi} {\bar u}_\chi(p_2) \Big((p_1+p_2)^\nu+i\sigma^{\nu\lambda}q_\lambda \Big) u_\chi(p_1),  \label{GI2}
\eea
we can rewrite 
\bea
16{\tilde T}^\chi_{\mu\nu} {\tilde T}^{\chi,\mu\nu}&=&\frac{(P\cdot K)}{2m^2_\chi}\, \bigg((K\cdot P)({\bar u}_\chi(k_2) u_\chi(k_1))({\bar u}_\chi(p_2)  u_\chi(p_1)) \nonumber \\
&&+ ({\bar u}_\chi(k_2) u_\chi(k_1))(K_\nu {\bar u}_\chi(p_2)i\sigma^{\nu\lambda} q_\lambda u_\chi(p_1))  \nonumber \\
&&-  (P_\mu{\bar u}_\chi(k_2)i\sigma^{\mu\rho}q_\rho u_\chi(k_1))({\bar u}_\chi(p_2)  u_\chi(p_1)) \nonumber \\
&&-( {\bar u}_\chi(k_2)i\sigma^{\mu\rho}q_\rho u_\chi(k_1))({\bar u}_\chi(p_2)i\sigma_{\mu\lambda} q^\lambda u_\chi(p_1))  \bigg)  \nonumber \\
&&- 4 m^2_\chi ({\bar u}_\chi(k_2) u_\chi(k_1))({\bar u}_\chi(p_2)  u_\chi(p_1))  \nonumber \\
&&+\frac{1}{2m^2_\chi} \bigg((K\cdot P)^2({\bar u}_\chi(k_2) u_\chi(k_1))({\bar u}_\chi(p_2)  u_\chi(p_1)) \nonumber \\
&&-(P_\mu {\bar u}_\chi(k_2)i\sigma^{\mu\rho}q_\rho u_\chi(k_1))(K_\nu {\bar u}_\chi(p_2)i\sigma^{\nu\lambda} q_\lambda u_\chi(p_1)) \nonumber \\
&&+(K\cdot P)({\bar u}_\chi(k_2) u_\chi(k_1))(K_\nu {\bar u}_\chi(p_2)i\sigma^{\nu\lambda} q_\lambda u_\chi(p_1)) \nonumber \\
&&-(K\cdot P)  (P_\mu {\bar u}_\chi(k_2)i\sigma^{\mu\rho}q_\rho u_\chi(k_1))({\bar u}_\chi(p_2)  u_\chi(p_1))\bigg) 
\label{tracefree}
\eea
where $P^\mu\equiv (p_1+p_2)^\mu$, $K^\mu\equiv (k_1+k_2)^\mu$ and $q^\mu\equiv (k_1-k_2)^\mu=(p_2-p_1)^\mu$.

On the other hand, the effective interactions for trace parts are
\bea
(T^\chi)^2 
= m^2_\chi  ({\bar u}_\chi(k_2)u_\chi(k_1))({\bar u}_\chi(p_2)u_\chi(p_1)). \label{trace}
\eea
Thus, the trace parts contain only scalar-scalar operators. 

Consequently, using eqs.~(\ref{tracefree}) and (\ref{trace}) and taking the non-relativistic limit and at small momentum transfer, we obtain the approximate self-scattering amplitude for fermion dark matter as
\bea
 {\cal M}_\chi&=&\frac{ic^2_\chi}{2\Lambda^2(m_G^2+{\vec q}^2) }\bigg[\bigg( \frac{1}{2}(P\cdot K)^2 +(P\cdot K){\vec q}^2  +\frac{1}{4}{\vec q}^4 -\frac{8}{3} m_\chi^4 \bigg)  {\mathcal{O}}_1^{\rm NR}\nonumber \\
&& -  m_\chi^2\Big(2 (P\cdot K)+{\vec q}^2\Big) {\mathcal{O}}_3^{\rm NR} - (P\cdot K){\vec q}^2  {\mathcal{O}}_4^{\rm NR}  \nonumber \\
&&- m_\chi^2 \Big( 2(P\cdot K)+ {\vec q}^2 \Big) {\mathcal{O}}_5^{\rm NR}+ m_\chi^2 (P\cdot K) {\mathcal{O}}_6^{\rm NR}+4 m_\chi^4  {\mathcal{O}}_3^{\rm NR} {\mathcal{O}}_5^{\rm NR} \bigg]  \nonumber \\
&\simeq  &\frac{ic_\chi^2}{2 \Lambda^2(m_G^2+{\vec q}^2)}\bigg[\bigg( \frac{16}{3} m_{\chi}^4  + 8 m_{\chi}^2 {\vec q}^2+8m^4_\chi  ({\vec v}^\bot)^2\bigg) {\mathcal{O}}_1^{\rm NR}  \nonumber \\
&&-m_\chi^2\Big(8 m_{\chi}^2 +3{\vec q}^2+4m^2_\chi ({\vec v}^\bot)^2 \Big)( {\mathcal{O}}_3^{\rm NR} +{\mathcal{O}}_5^{\rm NR} ) -4 m^2_\chi  {\vec q}^2  {\mathcal{O}}_4^{\rm NR} 
+4 m_\chi^4   {\mathcal{O}}_6^{\rm NR} \bigg] \label{feff}
\eea
where $P\equiv p_1+p_2$, $K=k_1+k_2$  and the non-relativistic operators used for direct detection \cite{DD} are given by
\bea
{\cal O}^{\rm NR}_1&=&1, \quad 
{\cal O}^{\rm NR}_2=(v^{\bot})^2, \quad 
{\cal O}^{\rm NR}_3= i{\vec s}_1\cdot \Big(\frac{{\vec q}}{m_\chi}\times {\vec v}^\bot\Big),  \nonumber \\
{\cal O}^{\rm NR}_4& =& {\vec s}_1 \cdot {\vec s}_2, \quad {\cal O}^{\rm NR}_5=i{\vec s}_2\cdot \Big(\frac{{\vec q}}{m_\chi}\times {\vec v}^\bot\Big), \quad {\cal O}^{\rm NR}_6 = \Big({\vec s}_1\cdot \frac{{\vec q}}{m_\chi}\Big)\Big({\vec s}_2\cdot \frac{{\vec q}}{m_\chi}\Big).
\eea
Here, we omitted the operator, $4  m_\chi^4  {\mathcal{O}}_3^{\rm NR} {\mathcal{O}}_5^{\rm NR}$, in the parenthesis of ${\cal M}_\chi$, because it is subdominant,  and used $2p_1\cdot k_1=s-2m^2_\chi=2p_2\cdot k_2$,  $2p_1\cdot k_2=-u+2m^2_\chi=2 p_2\cdot k_1$ for DM momenta,  $q^2\approx -{\vec q}^2$, and
\bea
P\cdot K=(p_1+p_2)\cdot (k_1+k_2)= s-u \simeq 4 m^2_\chi+2m^2_\chi v^2-{\vec q}^2.
\label{approx}
\eea
where use is made of  $s\simeq 4m_\chi^2(1+v^2/4)$ and $u= 4m^2_\chi-s-t\simeq -m^2_\chi v^2+{\vec q}^2$ in the non-relativistic limit with $v\equiv |{\vec v}|$.  We also used $v^2=({\vec v}^\bot)^2 +{\vec q}^2/m^2_\chi$ in the final result. 

\vspace{1cm}
\noindent
\underline {\bf Scalar dark matter}:
 
With eqs.~(\ref{scalar-emtensor}) and (\ref{scalar-trace}), the self-scattering amplitude for scalar dark matter is given by
\bea
{\cal M}_S&=& \frac{ic_S^2}{2 \Lambda^2(m^2_G-q^2)}  \left(2{\tilde T}^S_{\mu\nu} {\tilde T}^{S,\mu\nu} -\frac{1}{6} ({T}^S)^2 \right) \nonumber \\
&=&\frac{i c_S^2}{2\Lambda^2(m_G^2-q^2)}\bigg[4(k_1\cdot p_1)^2+4(k_1\cdot p_2)^2-\frac{8}{3}\Big( m_S^4-m_S^2(k_1 \cdot k_2)+ (k_1\cdot k_2)^2 \Big) \bigg] \nonumber \\
 &\simeq&\frac{i c_S^2}{2\Lambda^2(m_G^2+{\vec q}^2)}\bigg[ \frac{16}{3}m_S^4+\frac{8}{3}m_S^2\vec{q}^2+8m_S^4(\vec{v}^{\perp})^2  \bigg]
\label{self-scalar}
\eea
where we took the non-relativistic limit for scalar dark matter. 

\vspace{1cm}
\noindent
\underline {\bf Vector dark matter}: 

The self-scattering amplitude for vector dark matter, $\epsilon^{s\mu}(k_1)+\epsilon^{u\mu}(p_1)\rightarrow \epsilon^{s'\mu}(k_2)+\epsilon^{u'\mu}(p_2)$,  is given by
\bea
{\cal M}_X &=&  \frac{ic^2_X}{2\Lambda^2(m^2_G-q^2)} \left(2{\tilde T}^X_{\mu\nu} {\tilde T}^{X,\mu\nu} -\frac{1}{6} ({T}^X)^2 \right)  \nonumber \\
&=&{i c_X^2\over 2\Lambda^2(m_G^2-q^2)}\Big[ \eta_{\alpha\beta}\eta_{\rho\sigma}\Big({16\over 3}m_X^4+2s(s+t-4m_X^2)\Big)+t^2C_{\alpha\beta,\rho\sigma}  \nonumber\\
&&-2\eta_{\alpha\beta}\big( 2(t-m_X^2)p_{2\rho} p_{1\sigma} +(s+t-2m_X^2)(k_{2\rho} p_{1\sigma}+p_{2\rho} k_{1\sigma}) \nonumber \\
&&\hspace{1.5cm} +(s-2m_X^2)(k_{1\rho} p_{1\sigma} +p_{2\rho} k_{2\sigma})+ t(k_{2\rho} k_{1\sigma} + k_{1\rho} k_{2\sigma})  \big) \nonumber \\
&&-2\eta_{\rho\sigma}\big( 2(t-m_X^2)k_{2\alpha}k_{1\beta}+(s+t-2m_X^2)(p_{2\alpha}k_{1\beta}+k_{2\alpha}p_{1\beta}) \nonumber \\
&&\hspace{1.5cm} +(s-2m_X^2)(k_{2\alpha}p_{2\beta}+p_{1\alpha}k_{1\beta}) + t(p_{1\alpha}p_{2\beta}+p_{2\alpha}p_{1\beta})  \big) \nonumber \\
&&+2\eta_{\alpha\rho}\big( (s-2m_X^2)k_{1\beta}p_{1\sigma}+t(k_{1\beta}k_{2\sigma}+p_{2\beta}p_{1\sigma}) \big) \nonumber \\
&&+2\eta_{\alpha\sigma}\big( (s+t-2m_X^2)p_{2\rho}k_{1\beta}+t(p_{2\rho}p_{1\beta}+k_{2\rho}k_{1\beta}) \big)  \nonumber\\
&&+2\eta_{\beta\rho}\big( (s+t-2m_X^2)k_{2\alpha}p_{1\sigma}+t(k_{2\alpha}k_{1\sigma}+p_{2\alpha}p_{1\sigma}) \big)  \nonumber \\
&&+2\eta_{\beta\sigma}\big( (s-2m_X^2)k_{2\alpha}p_{2\rho}+t(p_{1\alpha}p_{2\rho}+k_{2\alpha}k_{1\rho}) \big)  \nonumber \\
&&+4k_{2\alpha}p_{1\beta}p_{2\rho}k_{1\sigma}+4p_{1\alpha}k_{1\beta}p_{2\rho}k_{2\sigma}+4p_{2\alpha}k_{1\beta}k_{2\rho}p_{1\sigma}+4k_{2\alpha}p_{2\beta}k_{1\rho}p_{1\sigma}  \nonumber \\
&&-8k_{2\alpha}k_{1\beta}p_{2\rho}p_{1\sigma}  \Big]\epsilon^\alpha(k_1)\epsilon^{*\beta}(k_2)\epsilon^\rho(p_1)\epsilon^{*\sigma}(p_2).
\eea
Then,  for $s\simeq 4m^2_X+ m^2_X v^2$ and $t=q^2\simeq -{\vec q}^2$ and keeping the leading terms in the momentum expansion, the above scattering amplitude becomes approximated to
\bea
{\cal M}_X&=&{i c_X^2\over \Lambda^2(m_G^2+{\vec q}^2)}\bigg\{\Big({8\over 3}m_X^4+4m^4_X (v^\perp)^2 \Big) {\cal S}^{s's}_{ii}{\cal S}^{r'r}_{ii}  \nonumber \\
&&+\frac{10}{3} m^3_X\, i ({\cal S}^{s's}_{ii}  {\vec S}^{r'r}_X+ {\cal S}^{r'r}_{ii}  {\vec S}^{s's}_X )\cdot ({\vec q}\times {\vec v}^\perp) \nonumber \\
&&+ \frac{2}{3}m^4_X ({\vec v}^\perp\cdot {\cal S}^{s's}\cdot {\vec v}^\perp)\, S^{r'r}_{ii} + \frac{2}{3} m^4_X({\vec v}^\perp\cdot {\cal S}^{r'r}\cdot {\vec v}^\perp)\, S^{s's}_{ii}  \nonumber \\
&& +2m^2_X \Big( -\frac{1}{3} {\cal S}^{s's}_{ii}\,({\vec q}\cdot {\cal S}^{r'r}\cdot {\vec q})-\frac{1}{3} {\cal S}^{r'r}_{ii}\,({\vec q}\cdot {\cal S}^{s's}\cdot {\vec q}) + {\cal S}^{rs}_{ii}\,({\vec q}\cdot {\cal S}^{r's'}\cdot {\vec q}) \nonumber \\
&&\quad+{\cal S}^{r's'}_{ii}\,({\vec q}\cdot {\cal S}^{rs}\cdot {\vec q})  - {\cal S}^{r's}_{ii}\,({\vec q}\cdot {\cal S}^{rs'}\cdot {\vec q})-{\cal S}^{rs'}_{ii}\,({\vec q}\cdot {\cal S}^{r's}\cdot {\vec q})\Big) \bigg\}. \label{VDM-M}
\eea
The above results with $ {\cal S}^{s's}_{ii}=\delta^{s's}$, ${\cal S}^{r'r}_{ii}=\delta^{r'r} $ and ${\vec S}^{s's}_X\cdot {\vec S}^{r'r}_X ={\cal S}^{s'r}_{ii}{\cal S}^{sr'}_{ii}-{\cal S}^{sr}_{ii}{\cal S}^{s'r'}_{ii}$ were used in the text.

\def\theequation{B.\arabic{equation}}

\setcounter{equation}{0}

\vskip0.8cm
\noindent
{\Large \bf Appendix B:  Non-relativistic operator basis} 
\vskip0.4cm
\noindent

For comparison, we note the following different notations for non-relativistic operators \cite{DD} shown in the appendix A and the text for fermion dark matter \cite{tanedo}, 
\bea
{\cal O}^{\rm NR}_1 &=&{\cal O}_1=1, \\
{\cal O}^{\rm NR}_4&=&{\cal O}_2={\vec s}_1\cdot {\vec s}_2,  \\
{\cal O}^{\rm NR}_6 &=& -{\cal O}_3=\frac{1}{m^2_\chi}\,({\vec s}_1\cdot {\vec q})({\vec s}_2\cdot {\vec q}),  \\
{\cal O}^{\rm NR}_3+{\cal O}^{\rm NR}_5&=& -{\cal O}_7=\frac{i}{m_\chi}\, ({\vec s}_1+{\vec s}_2)\cdot ({\vec q}\times {\vec v}^\perp), 
\eea
and
\bea
{\cal O}^{\rm NR}_3 {\cal O}^{\rm NR}_5 &=& -\frac{1}{m^2_\chi} ({\vec q}^2 (v^\perp)^2-({\vec q}\cdot {\vec v}^\perp)^2) \, {\cal O}_2  - (v^\perp)^2 {\cal O}_3+\frac{{\vec q}^2}{m^2_\chi}\, {\cal O}_4 -\frac{i{\vec q}\cdot {\vec v}^\perp}{m_\chi}\, {\cal O}_5
\eea
where
\bea
{\cal O}_4 &=& ({\vec s}_1\cdot{\vec v}^\perp) ({\vec s}_2\cdot{\vec v}^\perp), \\
{\cal O}_5 &=&-\frac{i}{m_\chi}\,\bigg[({\vec s}_1\cdot {\vec q}) ({\vec s}_2\cdot{\vec v}^\perp)+ ({\vec s}_2\cdot {\vec q})({\vec s}_1\cdot{\vec v}^\perp) \bigg].
\eea

\vskip0.8cm
\noindent
{\Large \bf Appendix C:  Momentum integral formulas for the effective potential } 
\vskip0.4cm
\noindent

The integral formula for the Yukawa-type potential is
 \bea
 \int \frac{d^3 q}{(2\pi)^3} \, e^{i{\vec q}\cdot {\vec r}}\, \frac{1}{{\vec q}^2+m^2_G} = \frac{e^{-m_G r}}{4\pi r}.
 \eea
 Then, for instance, for the effective scattering amplitude, ${\cal M}=i{\cal T}=\frac{4 iA^2_{\rm DM} m^2_\chi}{{\vec q}^2+m^2_G}$, the Yukawa potential can be obtained from the following Fourier transform,
 \bea
 V_{\rm eff} &=& -\frac{1}{4 m^2_\chi}  \int \frac{d^3 q}{(2\pi)^3} \, e^{i{\vec q}\cdot {\vec r}}\, {\cal T} \nonumber \\
 &=&- \frac{A^2_{\rm DM}}{4\pi r}\, e^{-m_G r}.
 \eea
Noting that the ${\vec q}$-dependent terms in the effective interactions correspond to the derivative terms in position space by
\bea
\nabla \, e^{i{\vec q}\cdot {\vec r}} &=& (i{\vec q})\, e^{i{\vec q}\cdot {\vec r}}, \\
\nabla^2 \, e^{i{\vec q}\cdot {\vec r}} &=& -{\vec q}^2\, e^{i{\vec q}\cdot {\vec r}},
\eea
we also get the useful formulas for the following integrals,
\bea
4\pi  \int \frac{d^3 q}{(2\pi)^3} \, e^{i{\vec q}\cdot {\vec r}}\, \frac{q_i}{{\vec q}^2+m^2_G}&=&-i\nabla_i\Big( \frac{e^{-m_G r}}{ r} \Big)= i\,{\hat r}_i\, \frac{e^{-m_G r}}{ r^2} \, (1+m_G r), \\
4\pi \int \frac{d^3 q}{(2\pi)^3} \, e^{i{\vec q}\cdot {\vec r}}\, \frac{{\vec q}^2}{{\vec q}^2+m^2_G}&=&-\nabla^2\Big( \frac{e^{-m_G r}}{ r} \Big)=  -\frac{m^2_G\, e^{-m_G r}}{ r} +4\pi \delta^3(r)\, e^{-m_G r},\\ 
  4\pi \int \frac{d^3 q}{(2\pi)^3} \, e^{i{\vec q}\cdot {\vec r}}\, \frac{{\vec q}^4}{{\vec q}^2+m^2_G}&=&(\nabla^2)^2\Big( \frac{e^{-m_G r}}{ r} \Big)=\frac{m^4_G\, e^{-m_G r}}{r},\\ 
  4\pi   \int \frac{d^3 q}{(2\pi)^3} \, e^{i{\vec q}\cdot {\vec r}}\, \frac{q_i q_j}{{\vec q}^2+m^2_G}&=&-\nabla_i \nabla_j \Big( \frac{e^{-m_G r}}{r} \Big)\nonumber \\
  &=&-\frac{e^{-m_G r}}{r^3}\bigg[(3 {\hat r}_i {\hat r}_j -\delta_{ij})\Big(1+m_Gr +\frac{1}{3} (m_Gr)^2 \Big) +\frac{1}{3}(m_G r)^2 \delta_{ij}\bigg] \nonumber \\
  &&+\frac{4\pi }{3}\, \delta^3(r) \, e^{-m_G r}\, \delta_{ij}.
\eea

\vskip0.8cm
\noindent
{\Large \bf Appendix D:  Sommerfeld factors} 
\vskip0.4cm
\noindent

With the Hulth\'en potential (\ref{hulthen}), the analytic solution for the radial wave function is given by
\bea
R_l= \frac{t^{l+1}}{z} \frac{e^{-iz/2}}{\Gamma(c)} \left|\frac{\Gamma(a)\Gamma(b)}{\Gamma(2iw)} \right| \,_2F_1(a,b,c;t) \label{radialw}
\eea 
where $z=2kr$, $w=\frac{k}{\delta}$  with $k=m_{\rm DM} v$ being the momentum of each particle, 
\bea
t&=& 1-e^{-z/(2w)},
\eea
and 
\bea
a&=& l+1+iw\Big( 1-\sqrt{1-x/w}\Big), \\
b&=& l+1+iw\Big( 1+\sqrt{1-x/w}\Big), \\
c&=& 2(l+1)
\eea
with $x=A_{\rm DM}/v$. 
For $r\ll 1/\delta$ or $t\ll 1$, the hypergeometric function becomes
\bea
_2F_1(a,b,c;t)\approx 1+ \frac{ab}{2c} \, t + {\cal O}(t^2), \quad\quad t\approx \frac{z}{2w}=\frac{kr}{w}. 
\eea
Therefore, the radial wave function  (\ref{radialw}) becomes in this limit 
\bea
R_l\approx \Big(\frac{1}{2w} \Big)^{l+1} z^l\cdot e^{-iz/2}\cdot \frac{1}{\Gamma(c)} \left|\frac{\Gamma(a)\Gamma(b)}{\Gamma(2iw)} \right|.  \label{radialw2}
\eea
As a consequence, the Sommerfeld factor for the partial wave cross section with angular momentum $l$ is given \cite{cassel} by
\bea
S_l = \left| \frac{(2l+1)!}{(l!)^2} \frac{\Gamma(2iw)(2w)^{l+1}}{\Gamma(l+1+2iw)}\,\frac{\partial^l R_l}{\partial z^l}\Bigg|_{z=0} \right|^2. 
\eea
Then, using eq.~(\ref{radialw2}), we obtain 
\bea
S_l &=& \left|\frac{\Gamma(a)\Gamma(b)}{\Gamma(l+1+2iw)}\frac{1}{l!} \right|^2, \label{Sommerfeld0}
\eea
which was used in the main text.

\end{document}